\documentclass[12pt]{aastex}
\usepackage{natbib}
\usepackage{threeparttable}
\usepackage{subfigure}
\usepackage{amssymb,amsmath}
\usepackage{hyperref}
\usepackage{graphicx}
\usepackage{epstopdf}
\usepackage{enumitem}
\usepackage{multirow}

\newcommand{\V}[1]{\mathbf{#1}}  

\newcommand\fsec{\mbox{$^{\rm{s}}_{^{\textnormal{\large .}}}$}}

\begin{document}
\pagestyle{plain}
\pagenumbering{arabic}

\title{VLA Measurements of Faraday Rotation through Coronal Mass Ejections}
\author{Jason E. Kooi$^{1,2}$, Patrick D. Fischer$^{2}$, Jacob J. Buffo$^{2}$, and Steven R. Spangler$^{2}$}
\affil{1) U.S. Naval Research Laboratory, Code 7213}
\affil{4555 Overlook Ave. SW, Washington, DC 20375, USA}
\affil{2) Department of Physics and Astronomy, University of Iowa}
\affil{Iowa City, IA, 52240, USA}
\email{jason.kooi@nrl.navy.mil}


\begin{abstract}
Coronal mass ejections (CMEs) are large-scale eruptions of plasma from the Sun that play an important role in space weather.  Faraday rotation (FR) is the rotation of the plane of polarization that results when a linearly polarized signal passes through a magnetized plasma such as a CME and is proportional to the path integral through the plasma of the electron density and the line of sight component of the magnetic field.  FR observations of a source near the Sun can provide information on the plasma structure of a CME shortly after launch; however, separating the contribution of the plasma density from the line of sight magnetic field is challenging.

We report on simultaneous white-light and radio observations made of three CMEs in August 2012.  We made sensitive Very Large Array (VLA) full-polarization observations using $1-2$ GHz frequencies of a ``constellation'' of radio sources through the solar corona at heliocentric distances that ranged from $6-15R_\odot$. Of the nine sources observed, three were occulted by CMEs: two sources (0842+1835 and 0900+1832) were occulted by a single CME and one source (0843+1547) was occulted by two CMEs.  In addition to our radioastronomical observations, which represent one of the first active hunts for CME Faraday rotation since \cite{Bird:1985} and the first active hunt using the VLA, we obtained white-light coronagraph images from the LASCO/C3 instrument to determine the Thomson scattering brightness, $\mathrm{B}_{\mathrm{T}}$, providing a means to independently estimate the plasma density and determine its contribution to the observed Faraday rotation.  

A constant density force-free flux rope embedded in the background corona was used to model the effects of the CMEs on $\mathrm{B}_{\mathrm{T}}$ and FR.  The single flux rope model successfully reproduces the observed $\mathrm{B}_{\mathrm{T}}$ and FR profiles for 0842+1835 and 0900+1832; however 0843+1547 was occulted by two CMEs.  Using the multiple viewpoints provided by LASCO/C3 and STEREO-A/COR2, we model observations of 0843+1547 using two flux ropes embedded in the background corona and demonstrate this model's ability to successfully reproduce both BT and FR profiles.  The plasma densities ($6-22\times10^3$ cm$^{-3}$) and axial magnetic field strengths ($2-12$ mG) inferred from our models are consistent with the modeling work of \cite{Liu:2007} and \cite{Jensen:2008}, as well as previous CME FR observations by \cite{Bird:1985}.
\end{abstract}




\section{Introduction}\label{sec:Intro}

Coronal mass ejections (CMEs) are large scale eruptions of plasma from the Sun that play an important role in space weather.  As technology continues to progress, the need for more reliable space weather predictions has increased.  The ejected material of a CME is associated with strong magnetic fields that can cause substantial geomagnetic storms at Earth \citep{Gosling:1991}.  The general picture of CME generation is as follows: magnetic field lines emerge through the convection zone of the dense photosphere and into the more tenuous plasma of the corona, generating bipolar magnetic regions \citep{Babcock:1961}.  The complex motions of the photosphere adjust and twist these field lines, strengthening them until some non-equilibrium state is reached.  After this, the magnetic energy is released and a CME erupts, carrying with it $10^{11}-10^{13}$ kg of magnetized plasma at speeds from hundreds to over 1000 km s$^{-1}$ \citep{Chen:2011}.

While CMEs have many shapes and sizes, the standard CME is generally characterized by a closed outer loop and typically has the so-called three-part structure: a bright outer loop, followed by a dark cavity which contains a bright core \citep{Illing&Hundhausen:1985}.  The bright outer loop is usually identified with the expelled coronal mass, the cavity with a flux rope, and the bright core with the erupted prominence.  Most other CME structures are believed to be a result of projection effects \citep{Schwenn:2006}.  When a normal CME is ejected near the center of the occulting disk of a coronagraph, it appears to surround the occulting disk, yielding what is known as a halo CME.  If it is partially off-center, it will have an apparent angular width between $120^\circ-360^\circ$, earning the title partial halo CME.  An exception to this is the narrow CME, which displays jet-like motions and is thought to be associated with open field line regions in the corona.  

Although CMEs have been an active field of research since their discovery in the 1970s \citep[e.g.,][]{MacQueen:1974,Gosling:1974,Brueckner:1974}, there is still much to understand.  While the plasma structure of a CME is typically modeled as a magnetic flux rope, there is no consensus on the effective trigger that initiates a CME.  Other issues include identifying what causes the shift towards non-equilibrium and how CMEs are accelerated after initiation \cite[see][for more details]{Chen:2011}.

Of particular importance for space weather considerations is the orientation of the CME magnetic field with respect to the geomagnetic field of the Earth.  Determining this orientation, however, is difficult.  
It has been well known that the vector photospheric magnetic field can be determined by Zeeman splitting of spectral lines.  Typical temperatures in CMEs range from $10^5 - 10^6$ K; consequently, the plasma is highly ionized and tenuous and therefore it is difficult to measure Zeeman splitting due to thermal and non-thermal broadening of the emission lines.  Spacecraft near the first Lagrangian point (L1) can measure local fields in situ, but these measurements would only allow a warning time of $\sim30$ minutes before arrival at the Earth \citep{Weimer:2002,Vogt:2006}.

Observations of Faraday rotation, the rotation of the plane of polarization of linearly polarized radiation as it propagates through a magnetized plasma, have been used for decades to determine the strength and structure of the coronal magnetic field and plasma density.  Beyond the quasi-static, small amplitude Faraday rotation observations characteristic of the coronal plasma, large amplitude transients associated with CMEs have also been detected and have the potential to improve understanding of CMEs.  Remote Faraday rotation measurements can also be performed on Earth, tracking a CME from initiation to at least $15R_\odot$.  Further, Faraday rotation provides information on the orientation of the CME's magnetic field with respect to the observer's line of sight (LOS) and can potentially be used to determine this orientation well before the CME reaches Earth \citep{Liu:2007}.  Finally, Faraday rotation observations of a source near the Sun could provide information on the plasma structure of a CME shortly after launch, potentially shedding light on the initiation process.


\subsection{Faraday Rotation}\label{sec:IntroductiontoFR}
This paper will deal with probing coronal mass ejections via Faraday rotation of radio waves from extragalactic radio sources.  Faraday rotation is a change in the polarization position angle $\chi$ of polarized radiation as it propagates through a magnetized plasma; this rotation in position angle, $\Delta \chi$, is given by 
\begin{equation}
\Delta \chi = \left[ \left( \frac{e^3}{2 \pi m_e^2 c^4}\right) \int_{LOS} n_e \mathbf{B} \cdot \mathbf{ds} \right] \lambda^2 = \left[\mathrm{RM}\right]\lambda^2\label{eq:FRintro}
\end{equation}
in cgs units.  In Equation~\eqref{eq:FRintro}, $n_e$ and $\mathbf{B}$ are the plasma electron density and vector magnetic field, respectively.  The fundamental physical constants in parenthesess, $e, m_e, \mbox{ and } c$ are, respectively, the fundamental charge, the mass of an electron, and the speed of light.  The term in parentheses has the numerical value $C_{FR} \equiv 2.631\times10^{-17}$ rad G$^{-1}$.  $\mathbf{ds}$ is an incremental vector representing the spatial increment along the line of sight (LOS), which is the path on which the radio waves propagate with positive $s$ is in the direction from the source to the observer. The subscript $LOS$ on the integral indicates an integral along the line of sight.  Finally, $\lambda$ indicates the wavelength of observation.  The term in square brackets is called the rotation measure (hereafter denoted by the variable RM and reported in the SI units of rad m$^{-2}$), and is the physical quantity retrieved in Faraday rotation measurements.

\begin{figure}[htbp]
	\begin{center}
	\includegraphics[height=2.5in, trim = {25mm 48mm 25mm 40mm}, clip]{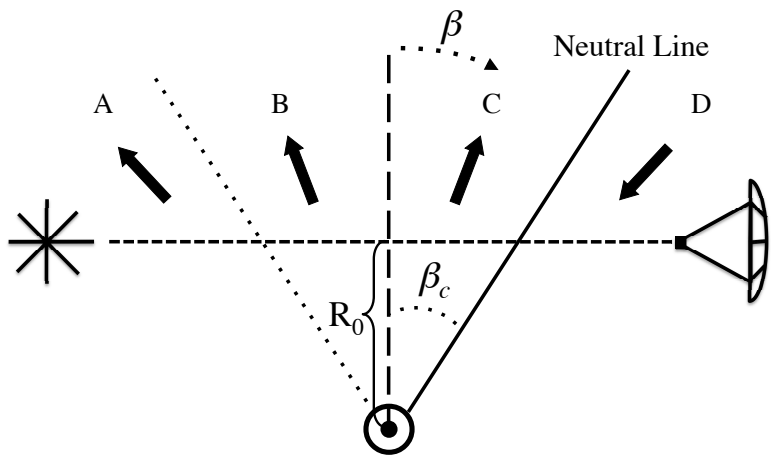}
	\caption[Illustration of the line of sight from a radio source, through the corona, to a radio telescope on Earth adapted to coronal conditions in 2012.]{\small Illustration of the line of sight from a radio source, through the corona, to a radio telescope on Earth.  The line of sight passes at a closest distance $R_0$ (impact parameter).  The figure illustrates an idealization that will be employed in this paper: the background coronal magnetic field is radial (solid arrows).  Faraday rotation depends sensitively on the location of the magnetic neutral line (solid line) along the line of sight (given by the angle $\beta_c$).  The dashed line divides the line of sight in two halves of equal length and the dotted line indicates a symmetry line.  This symmetry is such that, for the case of radial magnetic field strength dependent only on heliocentric distance and spherically symmetric plasma density, the RM contributions from sectors B and C ($-\beta_c<\beta<\beta_c$) cancel.  This figure is a modified version of Figure 1 in \cite{Kooi:2014} and has been adapted to coronal conditions in 2012 for 0846+1459 and 0843+1547.}
	\label{fig:cartoon1}
	\end{center}
\end{figure}

The geometry involved in a typical coronal Faraday rotation measurement is illustrated in Figure~\ref{fig:cartoon1}. The original version of this figure appeared in \cite{Kooi:2014}; however, Figure~\ref{fig:cartoon1} has been adapted to the coronal conditions in 2012 for the extragalactic radio sources 0846+1459 and 0843+1547, pertinent to the discussions that follow.  This figure illustrates two of the most important parameters for coronal Faraday rotation: the ``impact parameter,'' $R_0$, and the location of the magnetic neutral line along the line of sight, $\beta_c$.  The impact parameter is the shortest heliocentric distance of any point along the line of sight.  The neutral line gives the location at which the polarity of the vector magnetic field reverses and is usually associated with the coronal current sheet.  In Figure~\ref{fig:cartoon1}, the angle $\beta$ is used as an equivalent variable to $s$ for specifying the position along the line of sight and is defined as positive toward the observer.  

Faraday rotation measurements of the coronal plasma can be made with either spacecraft transmitters or natural radio sources as the source of radio waves.  Examples of results which have been obtained with spacecraft transmitters are \cite{Stelzried:1970}, \cite{Hollweg:1982}, \cite{Patzold:1987}, \cite{Bird&Edenhofer:1990}, \cite{Andreev:1997a}, \cite{Jensen:2013b,Jensen:2013a}, and \cite{Efimov:2015}.  Observations of natural radio sources typically use either pulsars or extragalactic sources.  Observations using pulsars include \cite{Bird:1980}, \cite{Ord:2007}, and \cite{You:2012} and observations utilizing extragalactic radio sources include \cite{Sofue:1972}, \cite{Soboleva&Timofeeva:1983}, \cite{Sakurai&Spangler:1994a,Sakurai&Spangler:1994b}, \cite{Mancuso&Spangler:1999,Mancuso&Spangler:2000}, \cite{Spangler:2005}, \cite{Ingleby:2007}, \cite{Mancuso&Garzelli:2013}, \cite{LeChat:2014}, and \cite{Kooi:2014}.  The advantage of using pulsars is that they can also be used to determine the dispersion delay through the corona simultaneously with Faraday rotation, thereby providing a means of independently estimating the plasma density contribution to the rotation measure. The corresponding disadvantage is that the dispersion delay due to the corona and solar wind is small and can not be measured with sufficient accuracy for most pulsars, with the exception of some millisecond pulsars \citep{You:2012}.


\subsection{White-light Imaging of CMEs}\label{sec:model_CME_whitelight_intro}

In studying coronal Faraday rotation, it is important to remember that Faraday rotation yields a path-integrated measurement of the magnetic field and the plasma density.  This presents two potential challenges.  The first is Faraday rotation measurements depend on the components of the magnetic field parallel (or antiparallel) to the line of sight; consequently, it is possible to measure zero Faraday rotation through the corona even when strong magnetic fields or dense plasmas are present.  A simple example would be a purely unipolar, radial magnetic field and $n_e = n_e(r)$ \citep[e.g., see][]{Kooi:2014}.  The second challenge is separating the contribution of the plasma density from the line of sight magnetic field to the RM.  The plasma density is typically determined independently from either models or observations.

In recent years, there have been considerable advances in space-based coronagraph technology, allowing for independent measurement of the coronal plasma density structure \citep[e.g., see][]{Jensen:2016}.  Modern observations of the corona are primarily obtained using white-light coronagraphs which observe radiation from the photosphere that has been Thomson-scattered by electrons in the coronal plasma.  The Thomson scattering brightness, $\mathrm{B}_{\mathrm{T}}$, is directly related to the coronal electron density, $n_e$, by the line of sight integral
\begin{equation}
\mathrm{B}_{\mathrm{T}} = \int_{LOS} n_e(\mathbf{r}) \mathcal{G}(\mathbf{r}) dr\label{eq:Thomsonintro}
\end{equation}
where $\mathbf{r}$ is vector heliocentric distance and $\mathcal{G}(\mathbf{r})$ is a geometric function determined by assumptions about solar limb darkening and heliocentric distance.  Provided the geometric function $\mathcal{G}(\mathbf{r})$ is known, the plasma density can be calculated by inverting Equation~\eqref{eq:Thomsonintro}.  A more detailed discussion of $\mathcal{G}(\mathbf{r})$ appears in Section~\ref{Sec:Model_background_ThomsonScattering}.

Over half a century ago, \cite{vandeHulst:1950} developed a method of deriving the coronal electron plasma density by inverting polarized brightness measurements.  Fifty years later, \cite{Hayes:2001} extended this technique to total brightness observations, allowing them to take full advantage of the extensive Large Angle and Spectrometric Coronagraph \citep[LASCO;][]{ Brueckner:1995} archive.  \cite{Hayes:2001} demonstrated that this total brightness technique yielded results as robust as the traditional methods of deriving coronal electron densities; further, total brightness measurements allow electron densities to be calculated at heights and in conditions inaccessible to polarized brightness observations.  However, because the electron corona (K corona) and scattering off interplanetary dust (F corona) both contribute to the total brightness, the accuracy of deriving $n_e$ from total brightness observations depends strongly on the accuracy of the removal of the brightness contributions from the F corona.

Total brightness imaging of the corona has also been applied to studying coronal mass ejections \citep{Vourlidas&Howard:2006}, providing information on many fundamental properties of coronal mass ejections, including mass \citep[e.g.,][]{Colaninno:2009}, speed and trajectory \citep[e.g.,][]{Morrill:2009}, and kinetic energy \citep[e.g.,][]{Vourlidas:2010}.   These total brightness techniques, originally developed for LASCO instruments on board the {\it Solar and Heliospheric Observatory} \citep[{\it SOHO};][]{Domingo:1995}, have also been extended to the Sun-Earth Connection Coronal and Heliospheric Investigation \citep[SECCHI;][]{Howard:2008} instrument suites on board the twin {\it Solar TErrestrial RElations Observatory} \citep[{\it STEREO};][]{Kaiser:2008} spacecraft.  Individually, the white-light imagers on {\it SOHO}, {\it STEREO-A}, and {\it STEREO-B} have been used to develop large online CME catalogs that employ both manual detection methods, such as those used in the {\it SOHO} LASCO CME Catalog \citep{Gopalswamy:2009}, as well as automated detection methods, as used in the Computer Aided CME Tracking software catalog \citep[CACTus;][]{Robbrecht:2009} or the Solar Eruptive Event Detection System \citep[SEEDS;][]{Olmedo:2008}.

However, the true power in these multiple white-light imaging instruments lies in combining measurements from {\it SOHO}, {\it STEREO-A}, and {\it STEREO-B}.  Using white-light measurement of CMEs from these multiple vantage points, \cite{Mierla:2010} reconstructed the three-dimensional structure of numerous, single CME events from 2007 and 2008.  \cite{Colaninno&Vourlidas:2015} similarly demonstrated the power of multiple-viewpoint observations by using  {\it SOHO}, {\it STEREO-A}, and {\it STEREO-B} to reconstruct the three-dimensional structures of three overlapping and interacting CMEs to shed insight into CME-CME interactions.  However, these observational advances still rely on models for the CME plasma structure to determine the electron plasma density; further these white-light observations can not provide direct measurements of the CME magnetic field structure.


\subsection{Previous Observations of CME Faraday Rotation}\label{sec:model_CME_FR_prevobser}

Most measurements of Faraday rotation transients caused by CMEs have been made by observing spacecraft transmitters.  Using {\it Pioneer 6}, \cite{Levy:1969} made the first measurements of Faraday rotation transients believed to be caused by coronal mass ejections in 1968 using an observational frequency of 2.292 GHz.  During these observations, the authors measured three large, ``W'' shaped transients 
at different heliocentric distances ($10.9R_\odot$, $8.6R_\odot$, and $6.2R_\odot$).  The transients were $\sim40^\circ$ (RM $\sim41$ rad m$^{-2}$) in amplitude and lasted for 2-3 hours and were each preceded by radio (decametric) noise-burst events.  \cite{Levy:1969} did not determine a definitive source for these transients, but concluded the events originated from a structure of significantly enhanced plasma density in the corona (i.e., they were not caused by ionospheric interference).

\cite{Cannon:1973} made similar measurements of Faraday rotation in the corona using {\it Pioneer 9}, again at a frequency of 2.292 GHz.  The authors 
observed two large transients.  The first (located $5.9R_\odot$ west of the Sun) had essentially the same $\sim40^\circ$ change in magnitude and the same negative rotation direction as the {\it Pioneer 6} observations
The second transient (located $6.2R_\odot$ east of the Sun) had a sigmoidal or inverse ``N'' shape, decreasing by $\sim7^\circ$ then increasing $\sim7^\circ$ above the steady-state rotation angle ($|\mathrm{RM}|\approx7.1$ rad m$^{-2}$) before leveling off over the course of five hours.  The authors found flares/subflares that coincided with both events and concluded that at least the first event probably resulted from the same type of phenomenon that caused the {\it Pioneer 6} transients. 

Whereas the previous measurements of coronal transients were fortuitous, \cite{Bird:1985} made a concerted effort to detect this phenomenon.  They used the {\it Solwind} coronagraph data to select intervals of Faraday rotation and spectral broadening measurements during solar occultations of  {\it Helios 1} and {\it Helios 2} during October and November of 1979.  They established a one-to-one correspondence between the five coronal transients observed in these two measurements and the passage of CMEs across the LOS.  To date, these observations are the highest quality observations of Faraday rotation anomalies due to CMEs.  Because these transients were very similar to the ones observed by \cite{Levy:1969}, \cite{Patzold:1998} concluded that CMEs were most likely responsible for the transients observed by \cite{Levy:1969} as well.

There is only one known measurement of a CME Faraday rotation anomaly during observations of an extragalactic radio source.  \cite{Spangler&Whiting:2009} indicated that the outer loop of a CME approached two sources (J2335-015 and J2337-025) during Faraday rotation observations at 1.465 GHz performed by \cite{Ingleby:2007}.  Although the LASCO/C2 coronagraph images suggested that the outer loop did not quite cross these LOS, the Faraday rotation of J2337-025 monotonously increased in time, increasing by $\sim26^\circ$ ($\mathrm{RM}\sim10.9$ rad m$^{-2}$) by the end of the observing session.


\subsection{Flux Rope Modeling of CMEs}\label{sec:model_CME_FR_intro}

One of the models that has become standard in describing CME morphologies is the flux rope.  In situ measurements of CMEs using several spacecraft (e.g., {\it Voyager 1}, {\it Voyager 2}, {\it Helios 1}, {\it Helios 2}, {\it IMP 8}, etc.) indicate that magnetic fields threading CMEs take the form of a helical flux rope \citep{Burlaga:1981,Burlaga:1988,Lepping:1990}.  This flux rope is either developed as part of the supporting structure necessary for the initial development of the solar prominence or as a result of field lines reconnecting during the eruption.  Aside from the spacecraft observations, the flux rope configuration also explains the white-light structure of CMEs \citep{Chen:1996,Gibson&Low:1998,Gibson&Low:2000}.  \cite{Gibson:2006} reported a more recent, comprehensive survey of white-light quiescent cavities (associated with a range of coronal loop morphologies) that suggested that the flux rope structure is formed prior to initiation of the CME.

Most models of CMEs describe the inner cavity as a flux rope \citep{Low:2001}; however, in forward modeling of CMEs captured by white-light imaging, the flux rope structure has also been used to describe the enhanced density structure of the bright outer loop preceding the inner cavity.  \cite{Thernisien:2006} modeled CMEs observed by the SECCHI/COR2 instruments on {\it STEREO-A} and {\it STEREO-B} using a Graduated Cylindrical Shell (GCS) flux rope structure in which the electrons are placed near the surface.  \cite{Wood:2009} similarly used a flux rope-like structure in modeling two distinct fronts of a CME on 2008 May 17.  Both \cite{Thernisien:2006} and \cite{Wood:2009} successfully reproduced the observed CME morphologies and determined electron densities (at the CME front) for these events.  More recently, \cite{Colaninno&Vourlidas:2015} used the GCS model to fit observations of three interacting CMEs and infer the trajectories, orientations, velocities, and source regions of these CMEs.

As mentioned previously, the orientation of the magnetic field is important in understanding the potential effects of a CME impacting the Earth's magnetosphere.  \cite{Liu:2007} have demonstrated that Faraday rotation measurements provide a remote-sensing method for determining this orientation well in advance of a CME's arrival at Earth.  They simulated Faraday rotation measurements using both force-free $\left(\nabla\times\V{B}=\alpha\V{B}\right)$ and non-force-free magnetic flux ropes and found that both types can (1) reproduce the signs and magnitudes of Faraday rotation transients previously associated with CMEs and (2) produce the same range in Faraday rotation profiles, from pseudo-Gaussian to ``N'' shaped profiles.  More importantly, the authors simulated a 2-D Faraday sky map of a flux rope CME approaching the Earth and argued that the full orientation and helicity of the CME could be remotely determined by Faraday rotation measurements using multiple LOS.

Building on this approach, \cite{Jensen:2008} attempted to reproduce the observational results of \cite{Levy:1969}, \cite{Cannon:1973}, and \cite{Bird:1985} using force-free flux ropes.  \cite{Jensen:2008} were able to reproduce the general ``V'' shape of the Faraday rotation profiles, but they could not reproduce the middle ``hump'' of the ``W'' shape of the {\it Pioneer 6} and {\it Helios} observations.  While they did not explore this discrepancy, \cite{Liu:2007} did note that two adjacent flux ropes with evolving fields could yield a ``W'' shaped profile.  Both \cite{Liu:2007} and \cite{Jensen:2008} found that multiple LOS are necessary for resolving any ambiguities in the magnetic field orientation or helicity.


\subsection{2012 Measurements of CME Faraday Rotation}
\label{sec:paper_outline}

In this paper, we present the results of observations of the radio galaxies 0842+1835, 0843+1547, and 0900+1832 which were occulted by coronal mass ejections on August 2, 2012. One of the advantages of using these extragalactic radio sources (relative to spacecraft transmitters and pulsars), which will be of importance in the present investigation, is that they are extended on the sky and, therefore, permit simultaneous measurement of Faraday rotation along as many lines of sight as there are source components with sufficiently large polarized intensities.  Obtaining this kind of information from spacecraft transmitters requires simultaneous tracking periods with two separated antennas \citep[see, e.g.,][]{Bird:2007}.  Another considerable advantage of extragalactic radio sources is that they emit, and are polarized, over a wide range in radio frequency, whereas spacecraft transmitters typically only provide one or two downlink frequencies.  Consequently, one can test for the $\lambda^2$ dependence of polarization position angle and resolve $n \pi$ ambiguities in the position angle ($n\in\mathbb{Z}$) and insure that the measured rotations in the position angle are indeed due to Faraday rotation.  

There are several reasons why these observations represent a significant improvement and extension of previous CME Faraday rotation experiments: 
\begin{enumerate}
\item These observations represent one of the first active hunts for CME Faraday rotation since \cite{Bird:1985} and is the first active hunt using the Very Large Array.
\item While several observations of satellite downlink signals have been made previously (generally at one frequency along one line of sight), these observations represent the first successful attempt to {\it actively} capture CME Faraday rotation with extragalactic radio sources, which provide multiple LOS over multiple frequencies.
\item These observations were made with the newly upgraded Very Large Array and, consequently, provide highly sensitive measurements of CME Faraday rotation.
\item Both 0842+1835 and 0900+1832 were occulted by one CME and 0843+1547 was occulted by two CMEs (one of which is the same CME that occulted 0842+1835), allowing for a strong test of the efficacy of flux rope models.
\item Unlike several previous studies \citep[e.g., ][]{Sakurai&Spangler:1994a,Sakurai&Spangler:1994b,Mancuso&Spangler:1999,Mancuso&Spangler:2000,Spangler:2005,Ingleby:2007,Kooi:2014}, we use simultaneous LASCO/C3 Thomson scattering data to independently determine the plasma density structure through the occulting CMEs. 
\item We observed in the B array configuration, and, therefore, are less susceptible to solar interference from the active regions near the solar limb that produced the occulting CMEs as well as the corresponding solar flares.
\item As a consequence of the previous points, the present observations are the most sensitive to date for our goal of measuring CME Faraday rotation and providing information for the CME plasma structure at heliocentric distances $\sim10R_\odot$.
\end{enumerate}

The organization of this paper is as follows.  In Section~\ref{sec:obs}, we discuss the source characteristics of radio galaxies 0842+1835, 0843+1547, 0846+1459, and 0900+1832, the geometry of the observations, the method for data reduction, and the imaging and analysis.    In Section~\ref{Sec:Model}, we discuss the models we employed for coronal Faraday rotation, Thomson scattering of white-light, and flux rope structure.  In Section~\ref{sec:results}, we present our results for the slow variations in rotation measure associated with the corona alone (0846+1459) and the rotation measure transients associated with occultation by CMEs (0842+1835, 0843+1547, and 0900+1832) as well as their associated model estimates for the plasma density and magnetic field structure of the occulting CMEs.  In Section~\ref{Sec:discussion}, we discuss the implications of our measurements  and compare our results with the observational results of \cite{Bird:1985} and modeling results of \cite{Liu:2007} and \cite{Jensen:2008}.  Finally, we summarize our results and conclusions in Section~\ref{sec:Summary}.


\section{Observations and Data Analysis}\label{sec:obs}
\subsection{Properties of the Target Radio Sources}\label{sec:source_properties}

\begin{figure}[htb!]
	\centering
	\includegraphics[height=4.75in, trim = {15mm 18mm 20mm 0mm}, clip]{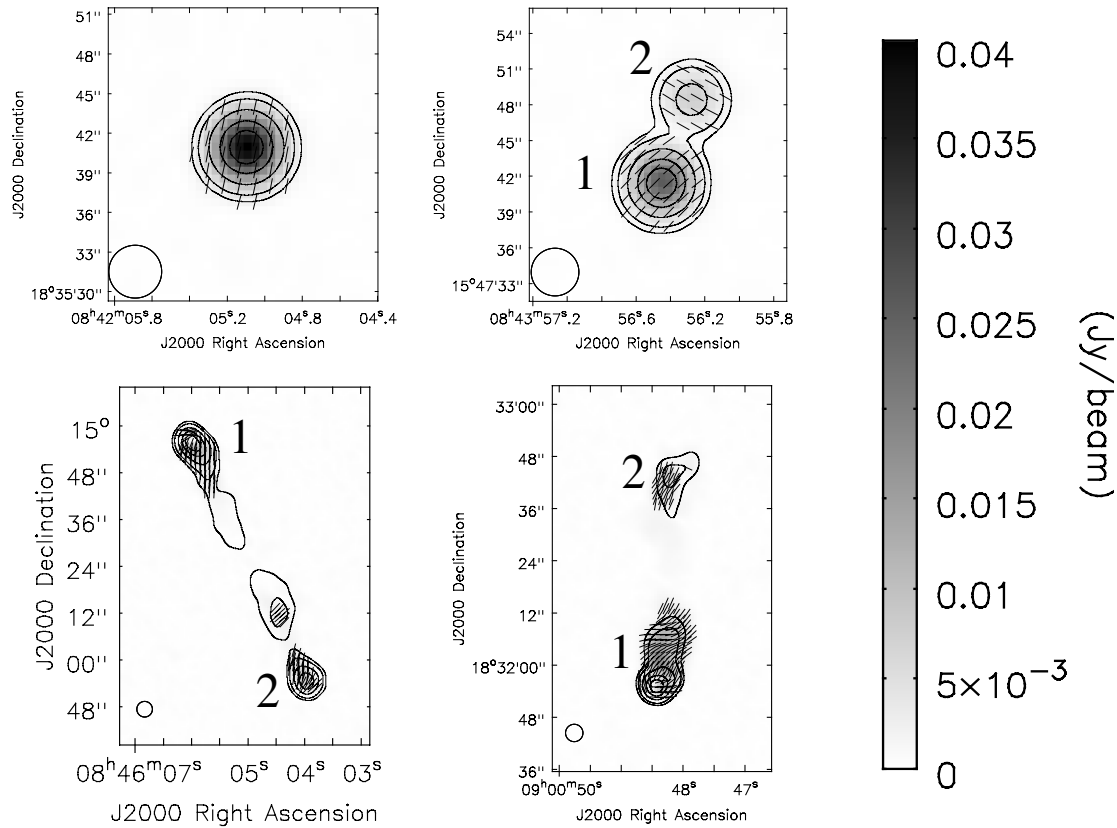}
%
%
	\caption[Clean map of the total intensity and polarization structure of the radio sources 0842, 0843, and 0846 on 2012 August 2.]
	 {\small Clean map of the total intensity and polarization structure of the radio sources 0842 ({\it top left}), 0843 ({\it top right}), 0846 ({\it bottom left}), and 0900 ({\it bottom right}) on 2012 August 30.  These images are a synthesis of the 56 MHz bandpass centered at a frequency of 1.845 GHz.  Contours show the distribution of total intensity (Stokes I), and are plotted at $-5$, 5, 10, 25, 50, and 75\% of the peak intensity for each source.
Grayscale indicates the magnitude of the polarized intensity.
Orientation of the line segments gives the polarization position angle $\chi$.  The labels ``1'' and ``2'' refer to Component 1 and Component 2, respectively, in the analysis presented in this paper.  The resolution of the image (FWHM diameter of the synthesized beam, plotted in the lower left corner of each image) is 4 arcseconds.}
	 \label{fig:Source_Maps}
\end{figure}

The basis of this paper is radioastronomical observations made in August, 2012, during the annual solar occultation of the extragalactic radio sources 0842+1835, 0843+1547, 0846+1459, and 0900+1832, henceforth referred to as 0842, 0843, 0846, and 0900 for the remainder this paper.
Images of the polarization structure of these sources made from our VLA observations when the Sun was far from the source (i.e., on the reference day, see Section~\ref{sec:Observation}) are shown in Figure~\ref{fig:Source_Maps} and details of the source characteristics are given in Table~\ref{T:source_character}.

\begin{table}[htbp]
	\centering
	\caption{Log of Observations for August, 2012}\label{T:source_character}
	\smallskip
	\begin{threeparttable}
		\begin{tabular}{@{\hspace{-6pt}}lllcccc}
		\hline\hline \noalign{\smallskip} 
		&\multicolumn{2}{l}{Dates of Observations}			&&	2012 Aug 2 			&	&  2012 Aug 30\\
		\hline\noalign{\smallskip}
		&\multicolumn{2}{l}{Duration of observing sessions (h)}	&	&	5.94 && 3.97\\
		&\multicolumn{2}{l}{Frequencies of observations (GHz)}	&	&	\multicolumn{3}{c}{$1.0-2.0$}\\
		&\multicolumn{2}{l}{VLA array configuration}			&	&	\multicolumn{3}{c}{B}\\
		&\multicolumn{2}{l}{Restoring beam (FWHM)\tnote{a}}		&&	\multicolumn{3}{c}{4\arcsec\tnote{b}}\\
		\noalign{\smallskip} 
		\hline\noalign{\smallskip} 
		&0842&Range in $R_0$ ($R_\odot$)					&	&	$9.6-10.6$		&&	$111.5-112.1$\\
		&& Component RA, DEC (J2000)				&&	\multicolumn{3}{c}{$08^{\rm{h}}42^{\rm{m}}95\fsec1+18\arcdeg35\arcmin41\arcsec$}\\
		&& $I$ (mJy beam$^{-1}$)\tnote{a,c}	&	&	$874\pm1.9$		&&	 $1070\pm0.4$\\
		&& $P$ (mJy beam$^{-1}$)\tnote{a,c}	 		&	&	$33.67\pm0.92$	&&	 $40.45\pm0.12$\\
		\noalign{\smallskip} 
		\hline\noalign{\smallskip} 
		&0843&Range in $R_0$ ($R_\odot$)					&	&	$9.9-10.5$		&&	$107.5-108.1$\\
		&& Component 1 RA, DEC (J2000)				&&	\multicolumn{3}{c}{$08^{\rm{h}}43^{\rm{m}}56\fsec5+15\arcdeg47\arcmin41\arcsec$}\\
		&& $I$ (mJy beam$^{-1}$)\tnote{a,c}	&	&	$327\pm0.5$		&&	 $375\pm0.3$\\
		&& $P$ (mJy beam$^{-1}$)\tnote{a,c}	 		&	&	$22.42\pm0.17$	&&	$25.95\pm0.12$ \\
		&& Component 2 RA, DEC (J2000)				&&	\multicolumn{3}{c}{$08^{\rm{h}}43^{\rm{m}}56\fsec3+15\arcdeg47\arcmin49\arcsec$}\\
		&& $I$ (mJy beam$^{-1}$)\tnote{a,c}				&	&	$109\pm0.5$		&&	 $126\pm0.3$\\
		&& $P$ (mJy beam$^{-1}$)\tnote{a,c}	 		&	&	$6.20\pm0.17$	&&	$7.65\pm0.12$ \\
		\noalign{\smallskip} 
		\hline\noalign{\smallskip} 
		&0846&Range in $R_0$ ($R_\odot$)					&	&	$11.1-11.4$			&&	$105.0-105.6$\\
		&& Component 1 RA, DEC (J2000)				&&	\multicolumn{3}{c}{$08^{\rm{h}}46^{\rm{m}}05\fsec9+14\arcdeg59\arcmin54\arcsec$}\\
		&& $I$ (mJy beam$^{-1}$)\tnote{a,c}	&	&	$65\pm0.5$			&&	 $70\pm0.5$\\
		&& $P$ (mJy beam$^{-1}$)\tnote{a,c}	 		&	&	$11.28\pm0.15$	&&	$11.92\pm0.14$ \\
		&& Component 2 RA, DEC (J2000)				&&	\multicolumn{3}{c}{$08^{\rm{h}}46^{\rm{m}}04\fsec0+14\arcdeg58\arcmin57\arcsec$}\\
		&& $I$ (mJy beam$^{-1}$)\tnote{a,c}				&	&	$90\pm0.5$		&&	 $100\pm0.5$\\
		&& $P$ (mJy beam$^{-1}$)\tnote{a,c}	 		&	&	$6.91\pm0.15$	&&	$7.40\pm0.14$ \\
		\noalign{\smallskip} 
		\hline\noalign{\smallskip} 
		\end{tabular}
	\end{threeparttable}
\end{table}
\begin{table}[htbp]
	\centering
	\smallskip
	\begin{threeparttable}
		\begin{tabular}{@{\hspace{-6pt}}lllcccc}
		\hline\noalign{\smallskip} 
		&0900&Range in $R_0$ ($R_\odot$)					&	&	$8.0-8.6$			&&	$95.4-96.0$\\
		&& Component 1 RA, DEC (J2000)				&&	\multicolumn{3}{c}{$09^{\rm{h}}00^{\rm{m}}48\fsec4+18\arcdeg32\arcmin01\arcsec$}\\
		&& $I$ (mJy beam$^{-1}$)\tnote{a,c}	&	&	$57\pm0.6$		&&	$67\pm0.6$\\
		&& $P$ (mJy beam$^{-1}$)\tnote{a,c}	 		&	&	$14.50\pm0.23$	&&	$18.48\pm0.14$ \\
		&& Component 2 RA, DEC (J2000)				&&	\multicolumn{3}{c}{$09^{\rm{h}}00^{\rm{m}}48\fsec2+18\arcdeg32\arcmin43\arcsec$}\\
		&& $I$ (mJy beam$^{-1}$)\tnote{a,c}				&	&	$28\pm0.6$		&&	 $36\pm0.6$\\
		&& $P$ (mJy beam$^{-1}$)\tnote{a,c}	 		&	&	$5.08\pm0.23$	&&	$7.12\pm0.14$ \\
		\noalign{\smallskip} 
		\hline
		\end{tabular}
	\medskip
		\begin{tablenotes}
		\footnotesize
		\item[a] This is for the maps using the data from the entire observation session.
		\item[b] The restoring beam on the day of occultation was fixed to be the same as on the reference day.
		\item[c] Mean and RMS levels for the 1.845 GHz maps (with bandwidth $\sim 56$ MHz).  
		\end{tablenotes}	
	\end{threeparttable}
\end{table}

0842 is a quasar and appears as a point source at our frequencies of observation.  While 0842 does not provide more than one line of sight, it is strongly polarized over these frequencies and so provides highly sensitive measurements.  The source 0843 is a radio source with two components and, consequently, provides two closely-spaced lines of sight through the corona: a strong central component and a weaker northern component (Components 1 and 2, respectively, in Table~\ref{T:source_character} and Figure~\ref{fig:Source_Maps}).  The source 0846 represents a distributed, polarized source of radio waves, ideal for probing of the corona with Faraday rotation, (e.g., see Figure~\ref{fig:Source_Maps}).  The polarized emission is strongest in the northern lobe and the southern lobe (Components 1 and 2, respectively, in Table~\ref{T:source_character} and Figure~\ref{fig:Source_Maps}), with two, much weaker components: a northern hotspot at (J2000) RA = $08^{\rm{h}}46^{\rm{m}}06\fsec1$ and DEC = $14\arcdeg59\arcmin58\arcsec$ and a southern jet at (J2000) RA = $08^{\rm{h}}46^{\rm{m}}04\fsec5$ and DEC = $14\arcdeg59\arcmin12\arcsec$.  The final source for discussion in this paper is 0900 and, like 0846, provides multiple LOS through the corona.  In this paper, we report the results for the strongest components in the southern lobe and the northern lobe, Components 1 and 2, respectively, in Table~\ref{T:source_character} and Figure~\ref{fig:Source_Maps}.

0843, 0846, and 0900 provide multiple lines of sight which pass through different parts of the corona and provide information on the spatial inhomogeneity of plasma density and magnetic field.  
The angular separations between the lines of sight to the northern and southern components of 0843, 0846, and 0900 are $7\farcs8$, $63\farcs7$, and $42\farcs1$, respectively, corresponding to 5,700 km, 46,000 km, and 31,000 km separation, respectively, between the lines of sight in the corona.

The peak intensity of 0846 was $I=100$ and $I=90$ mJy beam$^{-1}$ on the reference day and on the day of occultation by the corona, respectively.  Similarly, on the day of occultation, the peak polarized intensity decreases from 11.92 to 11.28 mJy beam$^{-1}$.  While the decrease in $I$ and $P$ for 0846 is minor on the day of occultation, there is a considerable decrease in peak $I$ and $P$ for 0842, 0843, and 0900 on the day of occultation.
The minor decrease in intensity of 0846 can be attributed to minor angular broadening effects typically associated with small scale coronal turbulence; however, both 0842, 0843, and 0900 were occulted by CMEs (Section~\ref{sec:Geometry}) and so the more substantial decreases in intensities of these three sources is probably due to angular broadening associated with these CMEs.  The effects of angular broadening are further discussed in Section~\ref{Sec:Imaging_Radio}.


\subsection{Properties of the Occulting CMEs}\label{sec:CME_properties}

\begin{figure}[htbp]
	\begin{center}
	\includegraphics[height=3.5in, trim = {10mm 0mm 0mm 0mm}, clip]{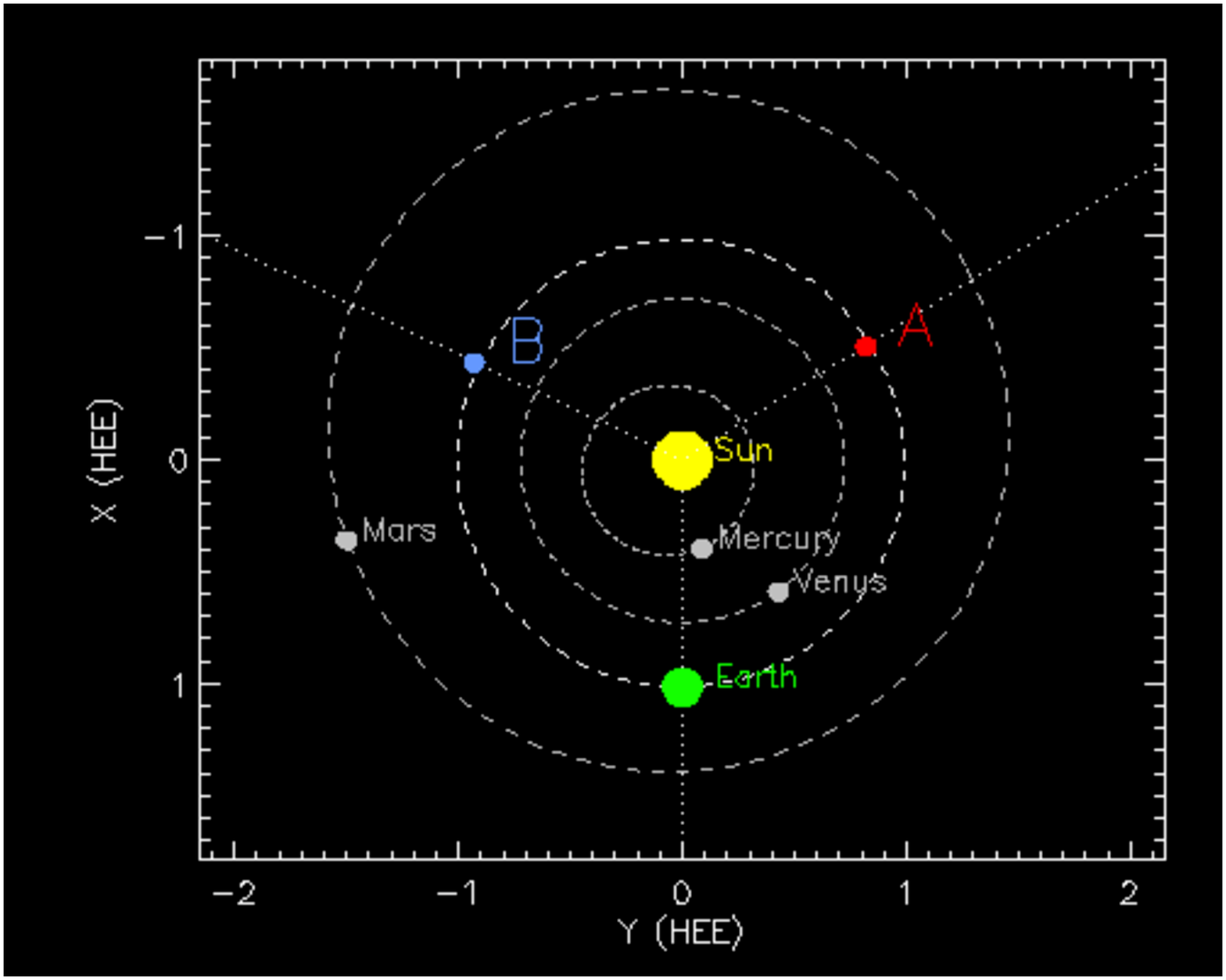}
	\caption[Positions of {\it STEREO-A} and {\it STEREO-B} relative to the Earth and {\it SOHO} at 18:00 UT on 2012 August 2.]{\small Positions of {\it STEREO-A} and {\it STEREO-B} relative to the Earth and {\it SOHO} at 18:00 UT on 2012 August 2.  {\it SOHO} is positioned near the Earth (green) at the L1 point, {\it STEREO-A} (red) is located $122\arcdeg$ ahead of the Earth, and {\it STEREO-B} (blue) is located $115\arcdeg$ behind the Earth.  The orbits of Mercury, Venus, and Mars are also shown for comparison.  The coordinates are Heliocentric Earth Ecliptic (HEE) and given in units of a.u.  This image was produced using the {\it STEREO} orbit tool online at \url{http://stereo-ssc.nascom.nasa.gov/}.}
	\label{fig:satellite_position}
	\end{center}
\end{figure}

Our total intensity white-light analysis in this paper relies primarily on coronagraph observations from the {\it SOHO} LASCO/C3 and the {\it STEREO-A} and {\it STEREO-B} COR2 instruments.  LASCO/C3 has a field of view (FOV) of $3.7-32R_\odot$, which overlaps the {\it STEREO-A} and {\it STEREO-B} COR2 field of view of $2.5-15R_\odot$, respectively.  The positions of {\it STEREO-A} and {\it STEREO-B} relative to the Earth and {\it SOHO} on the day of occultation (August 2, 2012) are given in Figure~\ref{fig:satellite_position}.
{\it SOHO} is positioned near the Earth at the L1 point and, on the day of occultation, {\it STEREO-A} was located $122\arcdeg$ ahead of the Earth (at a Carrington longitude and heliographic latitude of $284\fdg4$ and $0\fdg3$, respectively), and {\it STEREO-B} was located $115\arcdeg$ behind the Earth (at a Carrington longitude and heliographic latitude of $47\fdg6$ and $-6\fdg5$, respectively).   For comparison, the Carrington longitude and heliographic latitude of the center of the disk was $L0 = 162\fdg4$ and $B0 = 5\fdg9$, respectively.  Consequently, events appearing on the western limb of the Sun in LASCO/C3 appear just east of disk center in {\it STEREO-A} COR2 and events appearing on the eastern limb of the Sun in LASCO/C3 appear just west of disk center in {\it STEREO-B} COR2.  For the duration of this paper, we will refer to the COR2 instrument on {\it STEREO-A} and {\it STEREO-B} as COR2-A and COR2-B, respectively.

Data for all CMEs that occulted our radio sources appear in numerous CME catalogs; Table~\ref{T:CMEs_character} summarizes these data from three online catalogs: the {\it SOHO} LASCO CME Catalog \citep{Gopalswamy:2009}, Computer Aided CME Tracking software catalog \citep[CACTus;][]{Robbrecht:2009}, and the Solar Eruptive Event Detection System \citep[SEEDS;][]{Olmedo:2008}.  In this table, the position angle gives the orientation of the erupting CME and is measured counter-clockwise from Solar North; the angular width gives the approximate angular size  of the CME as measured from the sun; the linear velocity and acceleration are determined by fitting a first-order and second-order polynomial, respectively, to the height-time measurements for the event.

The first CME, henceforth referred to as CME-1, has the standard three-part structure described in Section~\ref{sec:Intro} and emerged from the southwestern limb of the Sun, entering the COR2-A and LASCO/C3 FOVs at 13:39 UT and 14:06 UT, respectively.
The emergence of CME-1 was coincident with the onset of a relatively weak solar flare (GOES Flare Class C1.5) that occurred near solar active region (NOAA \#) 11529.  The flare lasted from 12:10 UT to 13:35 UT and was visible at all wavelengths of the extreme-ultraviolet imager (EUVI) on {\it STEREO-A}.  This flare was located at a Carrington longitude and heliographic latitude of $249\fdg5$ and $-20$, respectively, which is within $5\arcdeg$ of the coronal magnetic neutral line; consequently, CME-1 originated within close proximity of the coronal magnetic neutral line.  This implies that CME-1 initiated near the solar limb on the Earth-side in LASCO/C3 images.  

The second CME, henceforth referred to as CME-2, also has the standard three-part structure and erupted from the southwestern limb of the Sun, entering the LASCO/C3 FOV at 15:54 UT.  While it appears almost two hours after CME-1 in LASCO/C3, it appears almost immediately after CME-1 in COR2-A, appearing at 13:54 UT.  The brightening feature in {\it STEREO-A} EUVI images due to the aforementioned C1.5 flare event travels north-west $20\arcdeg-30\arcdeg$ toward image center (the far-side in LASCO/C3 images), appearing to move along the coronal magnetic neutral line (the position of which was determined using data from the online archive of the Wilcox Solar Observatory, WSO, see Section~\ref{Sec:Model_background_Faradayrotation}).  
This brightening feature, which appeared in close proximity to the initiation point of CME-1 near 12:10 UT, finally disappears in close proximity to the initiation point of CME-2 near 13:35 UT (as determined by projecting the mean central position angle of the angles provided in column 4 of Table~\ref{T:CMEs_character} onto the photosphere).  Consequently, while CME catalogs such as the {\it SOHO} LASCO CME Catalog, CACTus, and SEEDS do not associate this CME with the C1.5 flare event, the location and timing of CME-2 suggest it is coincident with the conclusion of this flare.  For these reasons, we conclude that CME-2 erupted from the far-side of the sun, as seen in LASCO/C3 images, near a Carrington longitude and heliographic latitude of $\sim265\arcdeg$ and $\sim0\arcdeg$, respectively.

\begin{table}[htbp]
	\centering
	\caption{Occulting CME Characteristics on August 2, 2012}\label{T:CMEs_character}
	\smallskip
	\begin{threeparttable}
		\begin{tabular}{cccccccc}
		\hline\hline \noalign{\smallskip} 
		CME 	&	CME 			&&	Event	&	Position	&	Angular	&	Linear 			&	Acceleration\tnote{a}\\
		Identifier	&	Catalog			&&	Time		&	Angle	&	Width	&	Velocity			&	(m s$^{-2}$)	\\
				&					&&	(UT)		&	(deg)	&	(deg)	&	(km s$^{-1}$)		&	\\
		\noalign{\smallskip} 
		\hline\noalign{\smallskip}
				&	LASCO			&&	13:25	&	259		&	108	&	563			&	$-0.9$	\\
		CME-1	&	CACTus\tnote{b,c}	&&	$-$		&	$-$			&	$-$		&	$-$			&	$-$	\\
				&	SEEDS\tnote{b}	&&	13:25	&	247		&	84		&	491			&	$-0.1$	\\
		\noalign{\smallskip} 
		\hline\noalign{\smallskip} 
				&	LASCO			&&	14:48	&	286		&	120		&	412			&	$-1.5$	\\
		CME-2	&	CACTus\tnote{b}	&&	13:25	&	279		&	140		&	401			&	$-$	\\
				&	SEEDS\tnote{b}	&&	15:36	&	265		&	92		&	452			&	$-61.4$\tnote{d}	\\
		\noalign{\smallskip} 
		\hline\noalign{\smallskip} 
				&	LASCO			&&	16:36	&	47		&	26		&	649			&	$2.9$\tnote{e}	\\
		CME-3	&	CACTus\tnote{b}	&&	16:24	&	33		&	36		&	603			&	$-$	\\
				&	SEEDS\tnote{b}	&&	17:00	&	37		&	19		&	562			&	$23.5$	\\
		\noalign{\smallskip} 
		\hline
		\end{tabular}
	\medskip
		\begin{tablenotes}
		\footnotesize
		\item[a] The CACTus catalog does not provide acceleration estimates.
		\item[b] Both CACTus and SEEDS have LASCO-based and SECCHI-based catalogs; we report the values from the LASCO-based catalog for direct comparison to the {\it SOHO} LASCO CME Catalog.
		\item[c] There is signal confusion in both the LASCO-based and SECCHI-based CACTus catalogs between the CME-1 and CME-2 events because the CMEs overlap.
		\item[d] This value is likely a result of signal confusion between CME-1 and CME-2.
		\item[e] The {\it SOHO} LASCO CME Catalog notes that the acceleration is uncertain due to either (1) poor height measurement or (2) a small number of height-time measurements.
		\end{tablenotes}	
	\end{threeparttable}
\end{table}

The final CME, henceforth referred to as CME-3, emerged from the northeastern limb of the Sun, entering the COR2-B and LASCO/C3 FOVs at 16:09 UT and 16:54 UT, respectively.  CME-3 does not have an obvious three-part structure and has more in common with a narrow CME in that it displays a jet-like motion and arises near a pre-existing coronal streamer that is adjacent to a coronal dim region in LASCO/C3 images (e.g., see Figure~\ref{fig:CME_occult_geometry}).  This Thomson scattering dim region may be the consequence of a local coronal hole and, therefore, may be a region of unipolar flux (i.e., open magnetic field lines).  While the magnetic topology that may be inferred from LASCO/C3 images seems ideal for the production of a narrow CME, CME-3 has a larger angular width than typically defines narrow CMEs \citep[$<10\arcdeg$, see][]{Chen:2011} and there were no flare events near the northeastern limb of the Sun near the initiation time.  Because of this, we can not determine the point of eruption in solar coordinates as accurately as CME-1 and CME-2; however, from the position angles for CME-3 determined from COR2-B and LASCO/C3 images and from the location of the magnetic neutral line, we conclude that CME-3 initiated on the Earth-side of the Sun near a Carrington longitude and heliographic latitude of $\sim95\arcdeg$ and $\sim45\arcdeg$, respectively.  In calculating this, we assume that, like CME-1 and CME-2, CME-3 emerged in close proximity to the magnetic neutral line.


\subsection{Geometry of the Occultation}\label{sec:Geometry}

\begin{figure}[htb!]
	\begin{center}
	\subfigure[15:42 UT]{
	\includegraphics[height=2.2in, trim = {130mm 70mm 100mm 87mm}, clip]{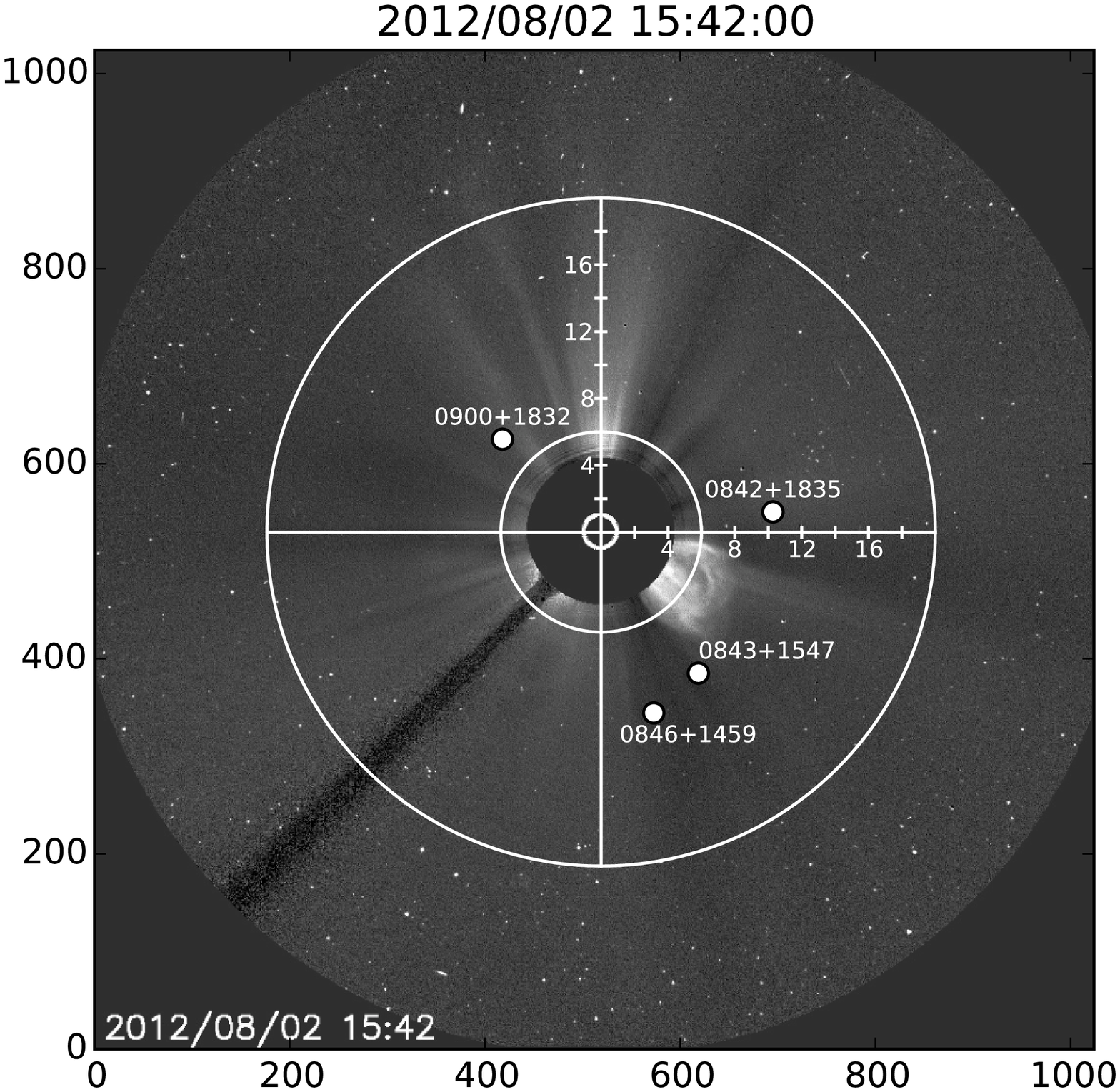}
	\label{fig:1542UTocc}
	}
	\hspace{.05in}
	\subfigure[17:42 UT]{
	\includegraphics[height=2.2in, trim = {130mm 70mm 100mm 87mm}, clip]{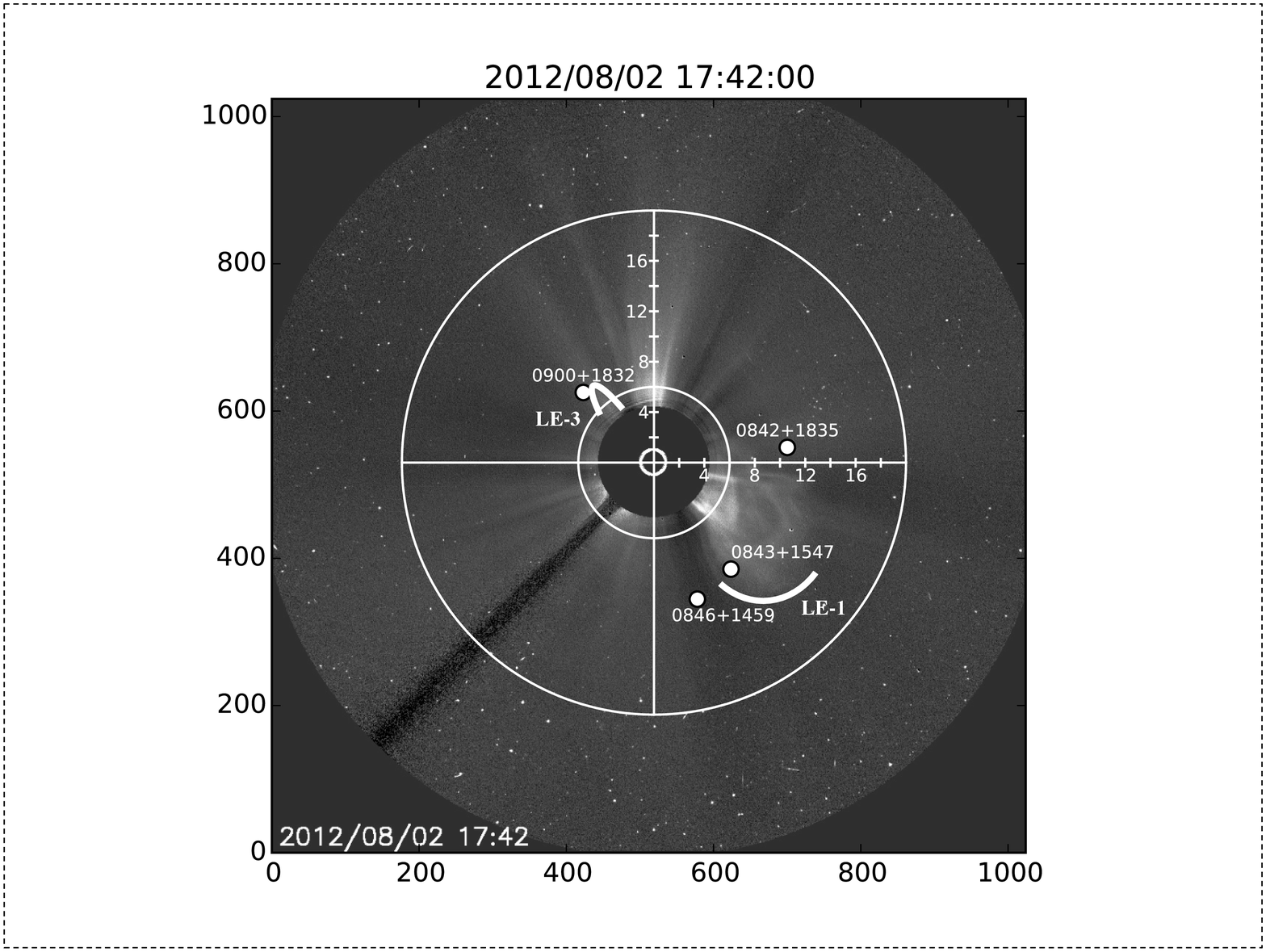}
	\label{fig:1742UTocc}
	}
         \vspace{.05in}
	\subfigure[19:06 UT]{
	\includegraphics[height=2.2in, trim = {130mm 70mm 100mm 87mm}, clip]{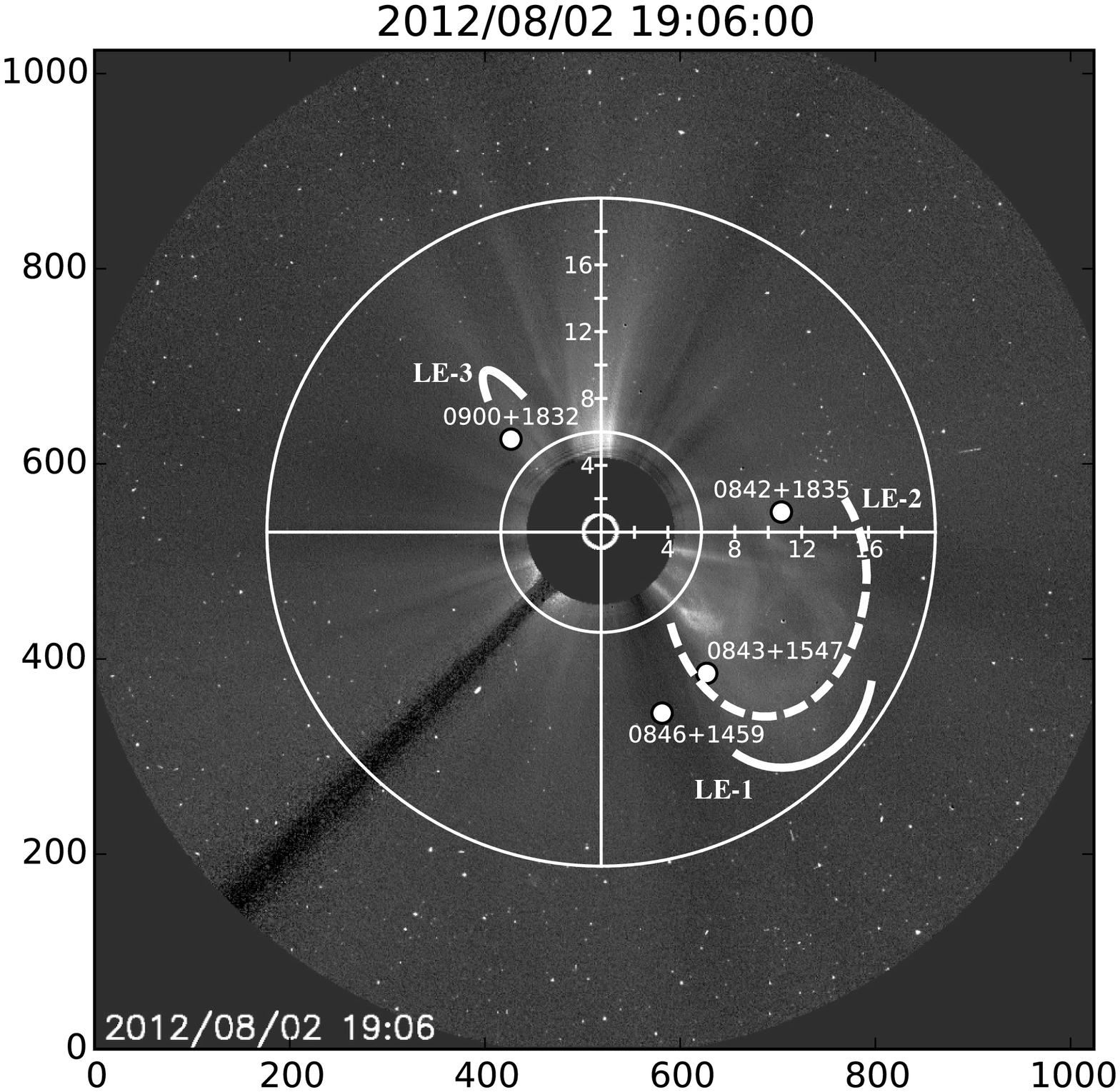}
	\label{fig:1906UTocc}
	}
	\hspace{.05in}
	\subfigure[20:30 UT]{
	\includegraphics[height=2.2in, trim = {130mm 70mm 100mm 87mm}, clip]{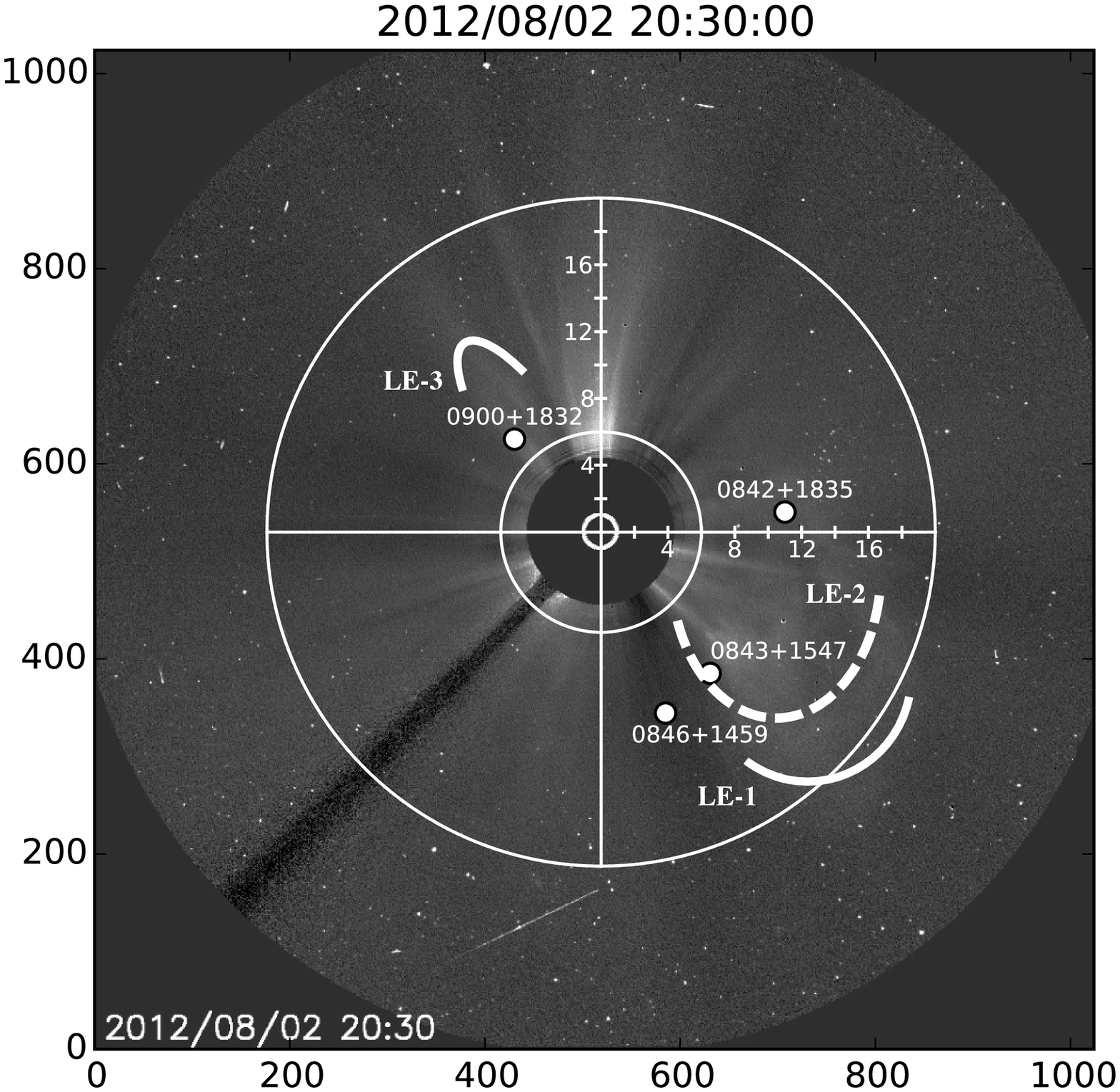}
	\label{fig:2030UTocc}
	}
	\caption[Corona and CMEs on 2012 August 2 observed with the LASCO/C3 coronagraph giving the geometry of the occultation.]{\small Corona and CMEs on 2012 August 2 as observed with the LASCO/C3 coronagraph.  White plotted points are the lines of sight to the radio sources (a) before occultation; (b) during occultation of 0842, 0843, and 0900 by CME-1, CME-2, and CME-3, respectively; (c) during occultation of 0843 by both CME-1 and CME-2; and (d) during occultation of 0843 by CME-2 only.  0846 was not occulted by a CME.  The solid curves (LE-1 and LE-3) and dashed curves (LE-2) represent the leading edges of CMEs originating on the Earth-side and far-side of the Sun, respectively.  The photosphere appears as the white circle centered inside the dark occulting disk and the horizontal axis is the heliographic equator with scale given in $R_\odot$.  Images are from the LASCO public archive: \url{http://sohowww.nascom.nasa.gov}.}	
	 \label{fig:CME_occult_geometry}
	\end{center}
\end{figure}

During the observing session, the orientations of the various lines of sight to our sources changed relative to the corona.  In performing coronal Faraday rotation observations, the most important parameter describing a given line of sight is the heliocentric distance to the proximate point along the line of sight, termed the impact parameter $R_0$.  The Carrington longitude and heliographic latitude of the proximate point are also important as they are used to determine the location where the line of sight crosses the coronal magnetic neutral line (the parameter $\beta_c$ in Figure 1).  During the August 2 session (details presented in Section~\ref{sec:Observation} below), the extended radio source 0846 was only occulted by the corona and was not occulted by a CME (Figure~\ref{fig:CME_occult_geometry}); the impact parameter ranged from $11.1-11.4R_\odot$ and there was a corresponding increase in the heliographic latitude of the proximate point from $-70\fdg4$ to $-66\fdg4$ and increase in the Carrington longitude from $224\fdg0$ to $225\fdg9$.

The radio source 0843 was slightly closer to the sun on the day of occultation, with a range in impact parameters of $9.9-10.5R_\odot$ corresponding to an increase in the heliographic latitude of the proximate point from $-50\fdg3$ to $-46\fdg6$ and decrease in the Carrington longitude from $240\fdg2$ to $238\fdg0$.  As may be seen in Figure~\ref{fig:CME_occult_geometry}, the line of sight to 0843 primarily sampled a coronal dim region prior to occultation; such regions may be associated with a coronal hole, where magnetic field lines are thought to be nearly unipolar and radial.

0843 was occulted by two CMEs on 2012 August 2: the first, CME-1, began occulting the line of sight to this source at 15:54 UT and the second, CME-2, began occultation at 18:30 UT.  By 20:06 UT, CME-1 passed beyond the LOS to 0843; however, CME-2 continued to occult this source until the end of the session.  Figure~\ref{fig:CME_occult_geometry} demonstrates the sequence of these events.  Because CME-1 and CME-2 overlap, we have outlined the leading edges of their bright outer loops (LE-1 and LE-2, respectively) in this figure.  From the vantage point of LASCO/C3, CME-1 is in the foreground and so the leading edge, LE-1, is denoted by a solid line in Figure~\ref{fig:CME_occult_geometry} and CME-2 is in the background and so the leading edge, LE-2, is denoted by a dashed line.

The quasar 0842 had the largest range in impact parameters ($9.6-10.6R_\odot$) because the LOS was located near the heliographic equator; the heliographic latitude of the proximate point decreased from $11\fdg8$ to $11\fdg2$ and the Carrington longitude decreased from $251\fdg0$ to $247\fdg4$.  As may be seen in Figure~\ref{fig:CME_occult_geometry}, 0842 was occulted by CME-2 beginning near 16:30 UT and continued to be occulted by this CME for the duration of observations.

The radio galaxy 0900 had the smallest impact parameters, ranging from $8.6R_\odot$ at the beginning of the session to $8.0R_\odot$ at the end, corresponding to an increase in the heliographic latitude of the proximate point from $38\fdg0$ to $42\fdg7$ and a decrease in the Carrington longitude from $72\fdg3$ to $68\fdg1$.  At the beginning of the observations, 0900 was occulted by a coronal streamer and, at 17:30 UT, the narrow, jet-like CME-3 began occulting this source and continued to do so for the remainder of the observing session.


\subsection{Observations and Data Reduction}\label{sec:Observation}

All radio observations were performed using the Karl G. Jansky Very Large Array (VLA) of the National Radio Astronomy Observatory (NRAO)\footnote{The Karl G. Jansky Very Large Array is an instrument of the National Radio Astronomy Observatory. The National Radio Astronomy Observatory is a facility of the National Science Foundation operated under cooperative agreement by Associated Universities, Inc.} and all data reduction was performed with the Common Astronomy Software Applications (CASA) data reduction package \citep{CASA:2007}.  Because CMEs can not currently be predicted with any precision, we had to make special arrangements with the staff at NRAO to schedule observations: we prepared a set of scheduling blocks for every day in the summer of 2012 (June 22 - August 20).

We selected a ``constellation'' of radio sources from the NRAO VLA Sky Survey \citep[NVSS;][]{Condon:1998} that would be occulted by the solar corona for each potential observation day.  We chose sources based on three primary criteria:
\begin{enumerate}
\item proximity to the Sun ($5-15R_\odot$)
\item degree of linear polarization ($P>5$ mJy beam$^{-1}$)
\item a requirement of $8-9$ of the strongest polarized sources evenly distributed around the Sun
\end{enumerate}
Because we would be observing at low frequencies ($1.0-2.0$ GHz), the Sun ($\sim 1$ MJy at 10 GHz) enters the sidelobes of the observing antennas at small impact parameters, increasing the noise in the signal considerably and preventing sensitive measurements typically $<5R_\odot$.  Beyond $15R_\odot$, coronal contributions to Faraday rotation are minimal and typically comparable in magnitude to ionospheric Faraday rotation.  Consequently, we chose sources within the range $5-15R_\odot$ where we were confident we could make sensitive measurements of CME-induced Faraday rotation.  Regarding the last point, we chose sources that were scattered around the Sun instead of a set of sources grouped in one region (e.g., aligned with an active region) because of the unpredictable nature of CMEs.  Having sources scattered around the Sun provides a better chance of measuring Faraday rotation through a CME, even if the CME only occults one or two sources. 

We closely monitored the Sun during these days and would submit a set of observations 24 hours in advance of the day on which we wished to observe.  We chose an observation day based on the following criteria:
\begin{enumerate}
\item multiple active regions were $\lesssim20\arcdeg$ of the solar limb
\item no major flare or CME events associated with these active regions in the previous 48 hours
\item increases in size of sunspots/sunspot groups associated with these active regions in the previous 48 hours
\item brightening in EUV images of these active regions as may be associated with strengthening magnetic fields in the previous 48 hours
\end{enumerate}
Regarding the first criterion, this increases the likelihood of capturing a CME demonstrating the traditional three-part structure in LASCO/C3 images.  The second criterion increases the likelihood of a large CME event because the solar active regions have not been releasing stored energy in recent CME events.  The other two criteria increase the probability of a CME erupting on the day of observation.  Based on these criteria, we made 3 sets of six-hour observations (August 2, August 5, and August 19) when sources were near the Sun and one set of four-hour reference observations (August 30) when all three sets of sources were distant from the Sun, allowing for measurement of the sources' intrinsic polarization properties, unmodified by the corona.  We were successful in capturing a CME on August 2.  Observations of the CME-occulted target sources performed on August 2, 2012, lasted from 14:46 to 20:53 UT and the reference observations performed on August 30, 2012, lasted from 15:04 to 19:03 UT.  While CMEs did erupt on August 5 and August 19, these CMEs did not emerge during the period of our observations on those days.

In this paper, we discuss the Faraday rotation to the three sources occulted by CMEs (0842, 0843, and 0900) as well as one source occulted only by the corona (0846) for comparison.  The details of these observations and resultant data are given in Table~\ref{T:source_character}.  Even though the other five sources from August 2 and the target sources from August 5 and 19 were not occulted by CMEs, these sources sample different regions of the corona -- with proximate points located at a range of heliographic latitudes and longitudes -- and provide further information on the global plasma structure of the corona at $5 - 15 R_\odot$.  Analysis and discussion of these data will appear in another paper in preparation.

The observations were similar in nature to those previously reported in \cite{Sakurai&Spangler:1994a}, \cite{Mancuso&Spangler:1999,Mancuso&Spangler:2000}, and \cite{Kooi:2014}, and described in those papers.  A detailed discussion of the main features of the observations are briefly summarized below.  We also indicate features of the 2012 observations which differ from those of our previous investigations. 

\begin{enumerate}
\item Observations were made in the B array.  This is important because we were purposely observing near strong solar active regions at the solar limb, which produce strong solar interference (i.e. strong, uneditable fringes due to active regions on the Sun) on short baselines.  In this configuration, the data are less affected by solar interference; however, the shortest interferometer baselines ($\leq4$ k$\lambda$) still had to be discarded.  This was done for all sources and for both sessions in order to allow more direct comparison between the reference day and the day of occultation.  In more compact array configurations, this restriction would represent a significant loss of data (e.g., the maximum UV distance at 1.845 GHz in the C and D configurations is 21 k$\lambda$ and 6.2 k$\lambda$, respectively, compared to 68 k$\lambda$ in B configuration).
\item We used an integration time of 15 seconds which, in B configuration, corresponds to an acceptable $\sim5\%$ time averaging loss in signal amplitude\footnote{See the Observational Status Summary documentation for the VLA at \url{https://science.nrao.edu/facilities/vla/docs}}.
\item Simultaneous observations were made at (L-band) frequencies of $1.0-2.0$ GHz divided into 16 bands, each with a resolution (channel width) of 1 MHz.
\item Due to radio frequency interference (RFI), large segments of the bandwidth had to be excised.  Of the original 16 frequency bands, we retained 7 with center frequencies of 1.356 GHz (bandwidth $=34$ MHz), 1.409 (bandwidth $=56$ MHz), 1.473 (bandwidth $=56$ MHz), 1.725 (bandwidth $=40$ MHz), 1.781 (bandwidth $=56$ MHz), 1.845 (bandwidth $=56$ MHz), and 1.899 (bandwidth $=37$ MHz).  Note the first three bands and, similarly, the other four bands are not contiguous; the edge channels of each band were removed.
\item Observations of the target sources were made in scans of $3-4$ minutes in duration; the time-on-source for a given scan depended on the magnitude of the polarized intensity, $P$, given in the NVSS catalog.  Each set of target scans was bracketed by 1.3 minute observations of a phase calibrator, for a total of 
10 scans on the day of occultation.  The average interval between each scan was $\sim32$ minutes.
\item The main calibrator for both sessions was J0825+0309.  This source was used for phase and amplitude calibration, as well as measurement of instrumental polarization.  In previous observations \citep[e.g.,][]{Kooi:2014}, we would observe a second calibrator source as an independent check of the polarimeter calibration.  We have always found that the polarization calibration values for both the primary phase calibrator and secondary phase calibrator were in excellent agreement; therefore, we did not include a second phase calibrator for these observations, choosing instead to maximize the time on our target sources.  The range in angular separation between this phase calibrator and the target sources was $12\fdg8-17\fdg6$.  These values are larger than typical VLA phase calibrator-target source separations ($\leq10\arcdeg$) because the phase calibrator needs to be far enough from the Sun to omit the possibility for the coronal plasma to influence this source.  On the day of occultation, the impact parameter for J0825+0309 was $R_0\sim60$; consequently, coronal influence on the calibration scans is negligible.  Further, previous investigations \citep[e.g.,][]{Ingleby:2007} performed sensitive Faraday rotation experiments with separations $\lesssim15\arcdeg$ and so our phase calibrator-target source separations are acceptable.
\item Polarization data were corrected for estimated ionospheric Faraday rotation using the Common Astronomy Software Applications (CASA) task \url{gencal} specifying the option \url{caltype = `tecim'}.  Prior to version 4.3.0, CASA did not have the ability to mitigate ionospheric Faraday rotation; however, George Moellenbrock (NRAO) and Jason Kooi implemented ionospheric Faraday rotation corrections in CASA version 4.3.0 and these corrections appear in all later versions \citep[see][]{Kooi:2016}.  The algorithm is similar to the Astronomical Image Processing System (AIPS) program procedure \url{VLBATECR} and functions by retrieving ionosphere model data from the Crustal Dynamics Data Information System (CDDIS), producing a CASA image file of the global vertical Total Electron Content (VTEC) values, and generates corrections for the ionospheric Faraday rotation based on these VTEC values.
Estimates for the ionospheric Faraday rotation measure ranged from $2.8-3.7$ rad m$^{-2}$ on August 2 and about $2.6-3.6$ rad m$^{-2}$ on August 30 for the four target sources.  The ionospheric Faraday rotation is similar for all target sources because they are within $5\arcdeg$ of each other; further, we observed at similar local sidereal times (LST) on both days.  Because of the method used to determine the coronal Faraday rotation (see Section~\ref{Sec:Imaging_Radio}), the total contribution from any residual ionospheric Faraday rotation should be negligible ($\leq0.1$ rad m$^{-2}$).
\item The instrumental polarization, described by the antenna-specific $D$ factors \\
\citep{Bignell:1982,Sakurai&Spangler:1994b} was determined from the observations of J0825+0309 in both sessions.  Even though we could not use the same reference antenna in both sessions, the amplitudes and phases of the $D$ factors were nearly identical for all antennas for both sessions.  Also, the amplitudes of the $D$ factors are higher for the upgraded VLA antennas, $D\approx 5-10\%$, than for the pre-upgrade antennas, $D\approx 1-4\%$, studied by \cite{Sakurai&Spangler:1994b}.  These results are similar to those in \cite{Kooi:2014}.
\item The net RL phase difference was determined using observations of 3C 286.  To test the precision of these calibration solutions, a second calibrator with known polarization, 3C 138, was calibrated using the RL phase difference solutions from 3C 286; the measured position angle was within $0\fdg3$ of the values listed in VLA calibrator catalogs for both sessions and all observing frequencies.
\item Of the 27 antennas used during these observations, three had to be excised on the reference day: two had abnormally high $D$ factors ($\sim50\%$) and one antenna's L-band receiver had been removed and, therefore, provided no data.  On the day of occultation, only one antenna had to be removed because its cross-hand (RL and LR) phases were poor after calibration.
\end{enumerate}


\subsection{Imaging with VLA Radio Data}\label{Sec:Imaging_Radio}

For each source, we generated maps in the Stokes parameters $I$, $Q$, $U$, and $V$ for each scan as well as a ``session map'' made from all the data on a given day, at a given frequency.  The session maps provide a measure of the mean Faraday rotation over the entire observing session; the individual scan maps, however, allow for examination of the time variations over the observing session, with a resolution on the order of the interval between scans: $\sim32$ minutes.

The imaging process was similar to the method described in \cite{Kooi:2014}; consequently, we will only indicate here features which differ from \cite{Kooi:2014}:
\begin{enumerate}
\item The calibrated VLA visibility data were split into the 7 bandpasses given in Section~\ref{sec:Observation} with center frequencies of 1.356 GHz (bandwidth $=34$ MHz), 1.409 (bandwidth $=56$ MHz), 1.473 (bandwidth $=56$ MHz), 1.725 (bandwidth $=40$ MHz), 1.781 (bandwidth $=56$ MHz), 1.845 (bandwidth $=56$ MHz), and 1.899 (bandwidth $=37$ MHz).  
\item Each bandpass was averaged in frequency from 1 MHz channel widths (resolution) to 4 MHz in order to expedite mapping.  We did not average over the whole bandpass because that would introduce significant bandwidth smearing effects.  
\item We used the CASA task \url{clean} using the multi-frequency synthesis mode with a cell size of $0\farcs6$ to generate the maps.  
\item We generated maps using a Natural Weighting scheme because we are primarily concerned with sensitivity and not resolution.
\item To accurately compare the maps restored at each of the 7 frequencies, the maps produced from observations on the day of occultation were restored using the same beam size ($4\farcs0$, the beam size for the lowest frequency bandpass) as maps from the reference observations on August 30; further, this same beam size was used to restore the maps at all frequencies.
\item One iteration of phase-only self-calibration was performed, which improved the ratio of peak intensity to the RMS noise (termed the dynamic range) by factors of $2 - 4$, depending on the bandpass.
\item We generated maps of the (linear) polarized intensity, $P$, and the polarization position angle, $\chi$, directly from the maps of Stokes $Q$ and $U$ according to $P = \sqrt{Q^2+U^2}$ and $\chi =0.5\arctan\left(U/Q\right)$.
\item We examined the session maps for local maxima in polarization intensity (typically $P>5$ mJy beam$^{-1}$) for each source on the reference day.  We chose these locations in order to maximize the sensitivity of our measurements because the error in measuring the polarization position angle is $\approx\sigma_P/2P$, where $\sigma_P$ is the error in measuring $P$; consequently, stronger $P$ provides a more robust measurement for $\chi$.  We then measured the values of the polarization quantities $I$, $Q$, and $U$ for the pixel with peak $P$ and derived the polarization quantities $P$ and $\chi$ for the individual scan and session maps on both observation days.
\item Of the extended sources, the polarized intensity of both the north and south components of 0843 are strong enough to allow accurate polarization measurements; however, the value of $P$ was too low to allow accurate measurements over most of the extended emission in 0846 and 0900.  In the rest of the paper, the analysis is based on measurements for the component in the northern and southern lobes where the polarized intensity was at a local maximum in the session map for the reference day, providing two lines of sight for both sources.
\item We calculated the coronal Faraday rotation, $\Delta\chi^i(\nu;x,y)$, for the $i$th scan map at frequency $\nu$ by straight subtraction:
\begin{eqnarray}\label{eq:RM_calc_diff}
\Delta\chi^i(\nu;x,y) &=& \chi^i_{occ}(\nu;x,y) - \chi_{ref}(\nu;x,y)
\end{eqnarray}
for $i\in[1,10]$, where $\chi^i_{occ}(\nu;x,y)$ are the polarization position angles at frequency $\nu$ and at location $(x,y)$ for the $i$th scan map on the day of occultation and $\chi_{ref}(\nu;x,y)$ is the polarization position angle for the same frequency and location on the session map for reference observations.  This subtraction method eliminates Faraday rotation caused by the background interstellar medium and typically reduces the effects of polarimeter calibration error, which would otherwise require second order instrumental polarization calibration \cite[e.g., see][]{Sakurai&Spangler:1994b}.
\item We then used a least squares algorithm to determine the rotation measure, RM, for each individual scan from the $\Delta\chi$ for each of the 7 bandpasses.  The fit is weighted by the radiometer noise because the fidelity of the data for the bandpasses centered at frequencies 1.409, 1.473, 1.781, and 1.845 GHz was superior to the other 3 bandpasses.
\end{enumerate}

Similar to \cite{Kooi:2014}, these maps demonstrate a lack of visible angular broadening of the radio sources; however, there is a measurable decrease in $I$ and $P$ on the day of occultation, particularly for the three sources occulted by CMEs.  In radioastronomical observations, the measured intensity is the convolution of the true intensity with a point spread function.  In these coronal observations, the point spread function for the target sources is the convolution of the synthesized beam with the power pattern of the angular broadening.  Following the procedure outlined in \cite{Kooi:2014}, we determined the Gaussian equivalent angular broadening disk and the corresponding drop in intensity for each target source.  

The angular broadening disk for the source that was occulted only by the corona, 0846, was asymmetric, but small ($1\farcs4\times0\farcs5$) and corresponds to a drop in intensity of $4-7\%$.  This is consistent with the decrease of $10\%$ and $5\%$ in $I$ and $P$, respectively, in Component 1.  The sources occulted by CMEs had more pronounced angular broadening associated with them:  the angular broadening disks for the extended sources, 0843 and 0900, were $1\farcs6\times0\farcs8$ and $3\farcs4\times1\farcs9$, respectively, corresponding to drops in intensity of $6-12\%$ and $13-16\%$, respectively.  Again, these are consistent with the decreases in peak $I$ and $P$ shown in Table~\ref{T:source_character} for these sources.  For 0842, the measured angular broadening disk is $1\farcs0\times0\farcs5$, corresponding to a decrease in intensity of $4\%$.  This is considerably less than the measured $18\%$ decrease for this source; however, it is difficult to measure angular broadening in 0842 because it is a point source and we have specified the restoring beam size.  Finally, as further evidence the phase calibrator was sufficiently far from coronal influences, there was no evidence of angular broadening for the phase calibrator.

The small size of the Gaussian disks for the target sources is also consistent with the lack of visible broadening in the maps (e.g., Figure~\ref{fig:Source_Maps}).  Because the broadening is not significant and is, in fact, smaller than the minimal effect measured in \cite{Kooi:2014}, we did not correct for this phenomenon (e.g., by convolving the session maps on the reference day with Gaussian equivalent disks).


\subsection{Imaging with LASCO/C3 White-Light Data}\label{Sec:Imaging_whitelight}

As discussed in Section~\ref{sec:model_CME_whitelight_intro}, observations of the corona are primarily obtained using white-light coronagraphs (e.g., LASCO/C3, COR2-A, and COR2-B) which observe radiation from the photosphere that has been Thomson-scattered by electrons in the coronal plasma.  In order to derive independent estimates for the plasma density, we use white-light images from the LASCO/C3 instrument.  LASCO/C3 is ideal because {\it SOHO} is aligned with the Earth and, therefore, the line of sight from a given radio source to the VLA in our radioastronomical data is similar to the the line of sight from that source to LASCO/C3 in optical white-light data.

Here, we outline the basic procedure we employed to produce LASCO/C3 images suitable for determining Thomson scattering brightness profiles for the lines of sight to each of our target sources:
\begin{enumerate}
\item We downloaded all LASCO/C3 Level 1 FITS images for August 2 as well as for the fifteen days prior to August 2 and the fifteen days following August 2, for a total of 31 days.  The Level 1 FITS images are scaled to the mean solar brightness, $B_\odot$, and have been pre-processed to correct for the flat field response of the detector, radiometric sensitivity, stray light, geometric distortion, and vignetting.  These images are made available to the public by the Naval Research Laboratory (NRL)\footnote{The {\it SOHO}/LASCO data used here are produced by a consortium of the Naval Research Laboratory (USA), Max-Planck-Institut fuer Aeronomie (Germany), Laboratoire d'Astronomie (France), and the University of Birmingham (UK).  {\it SOHO} is a project of international cooperation between ESA and NASA.} at \url{lasco-www.nrl.navy.mil}.
\item For each of the 31 days, we made a pixel-by-pixel median image.  By determining the median value pixel-by-pixel instead image-by-image, transients such as background stars, CMEs, and comets are removed.  We chose to make a pixel-by-pixel median image over a simpler pixel-by-pixel minimum image because a daily median image is less susceptible to particularly low brightness values as may be associated with a data gap due to interference or pre-processing issues.
\item We then produced the pixel-by-pixel minimum image for this 31-day period.  This final median-minimum image contained no signs of background stars, fast transients such as CMEs or comets, or slow transient structures such as coronal streamers; the final median-minimum image instead appears as a hazy, elliptical disk as may be associated with the F corona.
\item We subtracted the median-minimum image for the 31 day period from all LASCO/C3 Level 1 FITS images on August 2, the day of occultation.  The Thomson scattering brightness varies between $10^{-12}-10^{-11}B_\odot$ at heliocentric distances relevant to our radio observations ($8.0 - 11.4R_\odot$), which is consistent with model K corona brightness curves \citep[e.g., as presented in][]{Saito:1977,Hayes:2001}.
\item We developed Python code to determine the line of sight pixel position to all target sources in each LASCO/C3 subtraction image on the day of occultation.  The Thomson scattering brightness scaled to the mean solar brightness, $\mathrm{B}_{\mathrm{T}}/B_\odot$, for each line of sight was then given by the pixel value at this position.  Doing this for all LASCO/C3 subtraction images produces a Thomson scattering brightness time series for each target source with a time resolution of 12 minutes, the time interval between each LASCO/C3 image.  We only measure the Thomson brightness along one line of sight (to the target source center) even for the extended sources because the spatial resolution of the LASCO/C3 images is considerably lower (1 pixel $\sim0.06R_\odot$ $\sim13\farcm7$) than our radioastronomical observations ($4\farcs0$).
\item To estimate the the error in $\mathrm{B}_{\mathrm{T}}/B_\odot$, we calculated the mean value for $\mathrm{B}_{\mathrm{T}}/B_\odot$ in the outer field of view of the LASCO/C3 subtraction images (heliocentric distances of $25 - 30 R_\odot$), the region which is expected to be noise-dominated.  For all images, this value was within the range $0.2-0.5\times10^{-12}$.
\end{enumerate}

There are two issues that are important to consider in performing this median-minimum subtraction method.  First is the possibility that this method will not only remove the F corona, but will over subtract and remove a portion of the K corona contribution.  This is especially true if the K corona is quasi-static as is often the case during solar minimum conditions.  Our observations in 2012 were made during solar maximum and, therefore, the corona was very dynamic.  Over the 31 days used to produce the median-minimum F corona image, even large-scale quasi-static structures such as helmet streamers typically lasted less than $\lesssim5$ days.  Finally, we observed at heliocentric distances of $8.0 - 11.4R_\odot$; over-subtraction of the K corona is more pronounced at shorter distances.  For these reasons, it is unlikely that the Thomson scattering brightness time series for our target sources sample regions where the K corona is significantly over-subtracted.  The second consideration is whether we are noise-dominated.  Our Faraday rotation observations were already limited to $<20R_\odot$ because coronal Faraday rotation is negligible beyond this distance; consequently, our lines of sight are far from the outer FOV where the brightness is expected to be noise-dominated.

\begin{figure}[htbp]
	\begin{center}
	\includegraphics[height=3.35in, trim = {10mm 10mm 10mm 10mm}, clip]{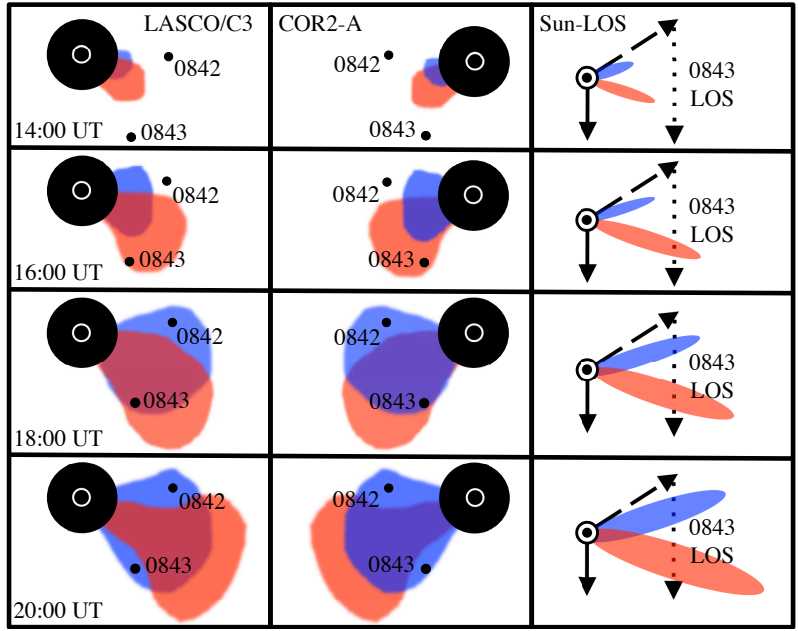}
	\caption[Illustration of the LASCO/C3 and COR2-A vantage points as well as a top-down view of the Sun-LOS plane.]{\small Illustration of the LASCO/C3 ({\it left column}) and COR2-A ({\it middle column}) vantage points as well as a top-down view of the Sun-LOS plane ({\it right column}).  The rows demonstrate the progression of CME-1 (red) and CME-2 (blue) before they occult 0843 near 14:00 UT ({\it top row}), during occultation by CME-1 alone near 16:00 UT ({\it second row}), during occultation by both CME-1 and CME-2 near 18:00 UT ({\it third row}), and during occultation by CME-2 alone near 20:00 UT ({\it bottom row}).  In the LASCO/C3 and COR2-A columns, the photosphere appears as the white circle centered inside the dark occulting disk, 0842 and 0843 appear as the black plotted points labeled ``0842'' and ``0843'', respectively, and the horizontal axis is the heliographic equator.  In the top-down view of the Sun-LOS plane for 0843, the solid arrow is directed toward LASCO/C3, the dashed arrow is directed toward COR2-A, and the dotted arrow gives the LOS to 0843.}
	\label{fig:cartoon_vantage_points}
	\end{center}
\end{figure}

The LASCO/C3 images also provide us with the unique ability to track the progression of the line of sight for a given target source through the CMEs observed on August 2 (e.g., Figure~\ref{fig:CME_occult_geometry}).  we measured the depth in pixels to which each line of sight penetrated a CME for every LASCO/C3 image.  This CME penetration depth, $y_p$, was used in modeling the Thomson brightness and Faraday rotation associated with occultation by a CME and is discussed in further detail in Section~\ref{Sec:Model_0842}.  For CMEs displaying the classical three-part structure (e.g., CME-1 and CME-2), the radius of the CME, $R_{cme}$, was also determined using these images.  For the analysis that follows, in measuring $R_{cme}$ we assumed that the flux tube consists of both the inner cavity and the outer loop and does not include the bright core.

Measuring $y_p$ and $R_{cme}$ for the line of sight to 0843 required additional information because it was occulted by both CME-1 and CME-2 from 18:30 UT to 20:06 UT.  To pinpoint the time at which 0843 was occulted by CME-2 and the time at which occultation by CME-1 ended, as well as to track the leading edges of CME-1 and CME-2 (LE-1 and LE-2, respectively), we relied on the additional vantage point provided by COR2-A.  Figure~\ref{fig:cartoon_vantage_points} gives an illustration of the LASCO/C3 and COR2-A vantage points as well as a top-down view of the Sun-LOS plane. The LASCO/C3 vantage point (left column) is similar to Figure~\ref{fig:CME_occult_geometry} with CME-1 (red) appearing in the foreground and CME-2 (blue) appearing in the background.  The geometry of {\it SOHO} and {\it STEREO-A} on the day of occultation is demonstrated in the right column of Figure~\ref{fig:cartoon_vantage_points} and, as a consequence of this geometry, CME-1 appears in the background of the illustration of the COR2-A vantage point (middle column) and CME-2 appears in the foreground.  The rows demonstrate the progression of CME-1 (red) and CME-2 (blue) as they occult 0843 during the observing session.

Tracing the approximate position of the line of sight to 0843 onto COR2-A images (downloaded from \url{http://secchi.nrl.navy.mil}) made it possible to follow the progression of this line of sight through CME-1 (appearing in the foreground of LASCO/C3 images and the background of COR2-A images) and CME-2 (appearing in the background of LASCO/C3 images and the foreground of COR2-A images).  Difference images (i.e., images produced by taking the pixel-by-pixel difference between the $i$th image and the $(i+1)$ image) for both LASCO/C3 and COR2-A were also used to more accurately track the leading edges LE-1 and LE-2.  It is worth noting that inclusion of this second CME is not merely adding more fit parameters, it is {\it required} by the independent {\it STEREO-A} data.  It is also important to emphasize that we could {\it not} perform the analysis that follows for 0843 without the multiple vantage points provided by the {\it SOHO} and {\it STEREO-A} satellites, as illustrated in Figure~\ref{fig:cartoon_vantage_points}.


\section{Coronal and Coronal Mass Ejection Models}\label{Sec:Model}

To obtain information on the plasma structure from Thomson scattering, Equation~\eqref{eq:Thomsonintro}, and Faraday rotation, Equation~\eqref{eq:FRintro}, we employ 
simplified analytic expressions for the plasma density and magnetic field.  We begin by first modeling the background coronal plasma; our ability to estimate the background corona is important for correctly interpreting the CME data.  We then employ flux rope models to reproduce observations of CME-1, CME-2, and CME-3.


\subsection{Modeling the Background Corona}\label{Sec:Model_background_corona}

\subsubsection{Model for Coronal Thomson Scattering}\label{Sec:Model_background_ThomsonScattering}

In our model for the background coronal plasma, we assume that the plasma density depends only on the heliocentric distance, $r$.  To model the coronal Thomson scattering brightness, $\mathrm{B}_{\mathrm{T}}$, we must determine the form of the geometric function $\mathcal{G}(\mathbf{r})$ in Equation~\eqref{eq:Thomsonintro}.  As discussed in Section~\ref{sec:model_CME_whitelight_intro}, $\mathcal{G}(\mathbf{r})$ depends on assumptions about solar limb darkening and heliocentric distance.  For spherically symmetric plasma, $\mathcal{G}(\mathbf{r})=\mathcal{G}(r)$.  The full form of $\mathcal{G}(r)$ \citep[given as Equation (17) in ][]{vandeHulst:1950} is
\begin{equation}
\mathcal{G}(r) = \left(\frac{3}{4}\sigma_T R_\odot B_\odot\right)\left[\left(2-\frac{R_0^2}{r^2}\right)\mathcal{A}(r)+ \frac{R_0^2}{r^2}\mathcal{B}(r)\right]\frac{r}{\sqrt{r^2-R^2_0}} \label{eq:geometric_function}
\end{equation}
where $\sigma_T$, $R_\odot$, and $B_\odot$, are the Thomson scattering cross-section, solar radius, and mean surface brightness of the Sun. $r$ is the heliocentric distance to a given point along the line of sight at which scattering occurs and $R_0$ is the impact parameter for the line of sight, both given here in units of $R_\odot$. $\mathcal{A}(r)$ and $\mathcal{B}(r)$ are geometric factors such that $\mathcal{A}(r)$ is the fraction of $2\mathcal{A}(r)+\mathcal{B}(r)$ that is proportional to the mean square of the electric field vector components in any transversal direction and $\mathcal{B}(r)$ is the fraction of $2\mathcal{A}(r)+\mathcal{B}(r)$ that is proportional to the mean square of the vector components in the radial direction.  The functional forms of $\mathcal{A}(r)$ and $\mathcal{B}(r)$ and the geometry involved in coronal Thomson scattering are given in \cite{vandeHulst:1950} and are not repeated here.

We treat the Sun as a point source in Equation~\eqref{eq:geometric_function}.  In this limit, $\mathcal{A}(r)\rightarrow1/2 r^{-2}$ and $\mathcal{B}(r)\rightarrow0$.  As demonstrated in \cite{vandeHulst:1950}, $\mathcal{A}(r)$ and $\mathcal{B}(r)$ rapidly approach these limits, reaching them by heliocentric distances of $5R_\odot$.  We are interested in Thomson scattering at impact parameters $\geq8R_\odot$ (see Table~\ref{T:source_character}); consequently, this assumption is valid for our purposes.  Applying this assumption and redefining Equations~\eqref{eq:Thomsonintro} and~\eqref{eq:geometric_function} in terms of the $\beta$ angle defined in Figure 1 gives the form
\begin{equation}
\mathrm{B}_{\mathrm{T}}/B_\odot = \left(\frac{3\sigma_T R_\odot}{16R_0}\right)\int^{\pi/2}_{-\pi/2}\left[1+\sin^2(\beta)\right]n_e(R_0,\beta)d\beta \label{eq:thomson_scattering_general}
\end{equation}

The specific form for the plasma density we choose is a single power law representation:
\begin{eqnarray}
n_e(r) = N_0 r^{-\alpha} \label{eq:single_nepower}
\end{eqnarray} 
where $N_0$ and $\alpha$ are free parameters and $r$ is in units of $R_\odot$.  The resulting expression for Thomson scattering is then
\begin{equation}
\mathrm{B}_{\mathrm{T}}/B_\odot = \left(\frac{3\sigma_T R_\odot N_0}{8}\right)R^{-\alpha-1}_0\left(\frac{\sqrt{\pi}}{1+\alpha}\right)\frac{\Gamma\left(\frac{5}{2}+\frac{\alpha}{2}\right)}{\Gamma\left(2+\frac{\alpha}{2}\right)} \label{eq:thomson_scattering_powerlaw}
\end{equation}
In particular, we use the same model value $\alpha=2.36$ as \cite{Kooi:2014}.  This power law gives predictions that have been in fairly good agreement with measurements reported in \cite{Sakurai&Spangler:1994a}, \cite{Spangler:2005}, and \cite{Ingleby:2007}.  While there have been a number of alternative power laws presented over the years \citep[e.g., $N_0 = 1.61 \times 10^6$ cm$^{-3}$ and $\alpha=2.45$ in][]{Patzold:1987}, the exact form of the power laws assumed in Equation~\eqref{eq:single_nepower} should not be crucial for the results presented here for two reasons.  First, the different functional forms give very similar values at heliocentric distances characteristic of our observations and, second, our observations were made in a narrow range of impact parameters ($8-11.4 R_\odot$).

We specify $\alpha$ and determine $N_0$ by fitting Equation~\eqref{eq:thomson_scattering_powerlaw} to the Thomson scattering profile for a given source using a least-squares method.  For 0846, we fit to the $\mathrm{B}_{\mathrm{T}}$ data over the entire observing period because the source was not occulted by a CME.  For sources occulted by a CME during radio observations, we fit to the $\mathrm{B}_{\mathrm{T}}$ data for the three hours prior to occultation by the leading edge of the CME.  The values of $N_0$ determined from each fit for a given line of sight are shown in Table~\ref{T:background_model_parameters} and the corresponding $\mathrm{B}_{\mathrm{T}}$ curve for the background coronal plasma is given as 
as a dotted line in the Thomson scattering brightness profile of Figures~\ref{fig:RM_Time_Series_0842_withmodel},~\ref{fig:RM_Time_Series_0900_withmodel}, and~\ref{fig:RM_Time_Series_0843_withdoublemodel} here in Section~\ref{Sec:Model}.  Discussion of the significance of the comparison of data and model is deferred to Section~\ref{sec:results} below.


\subsubsection{Model for Coronal Faraday Rotation}\label{Sec:Model_background_Faradayrotation}

In our model for the background coronal plasma, we assume that the plasma density depends only on the heliocentric distance, $r$, and that the magnetic field is entirely radial with its magnitude depending solely on $r$.  Frequently, the coronal magnetic field is approximated using some form of the {\it Dipole plus Current Sheet} (DCS) magnetic field \citep{Gleeson&Axford:1976} -- sometimes called a split monopole because of its topology -- or the {\it Dipole plus Quadrupole plus Current Sheet} (DQCS) model of \cite{Banaszkiewicz:1998} which adds a weak quadrupole term to the DCS model.  At these heliocentric distances, though, a radial magnetic field is a good approximation \citep[see, e.g.,][]{Banaszkiewicz:1998}.  However, we do retain the coronal current sheet of the DCS model as an infinitely thin neutral line, where the polarity of the coronal magnetic field reverses, located at an angle $\beta_c$.  This geometry is demonstrated in Figure 1.

As discussed in Section~\ref{sec:IntroductiontoFR}, the angle $\beta_c$ is crucial; the magnitude of the observed rotation measure is critically dependent on this parameter.  To determine $\beta_c$, we used the same procedure outlined in \cite{Kooi:2014}: a python program was used to project the line of sight for a given source onto heliographic coordinates.  Maps of the coronal magnetic field (determined by a potential field source surface model with the surface at $r=3.25 R_{\odot}$) were obtained from the online archive of the Wilcox Solar Observatory (WSO).  The digital form of these maps was used to determine the heliographic coordinates of the coronal neutral line.  The value of $\beta$ at which these two curves intersected gave the parameter $\beta_c$.  For further details, see \cite{Mancuso&Spangler:2000} and \cite{Ingleby:2007}.  

Because our observations were made during solar maximum conditions, the neutral line has a complex geometry and crosses the lines of sight for several sources from our August, 2012 observations multiple times; however, for 0842, 0843, and 0846, the associated lines of sight only cross one neutral line as shown in Figure 1.  Under these symmetric conditions, the contributions to the integral in Equation (1) 
from zones B \& C cancel, while those of A \& D make equal contributions of the same sign.  The line of sight to 0900, though, crosses two neutral lines and this second crossing must be accounted for to properly model the background coronal Faraday rotation to this source.

We use the same form for the coronal plasma density as Equation~\eqref{eq:single_nepower}, with $\alpha=2.36$ and $N_0$ determined from the least-squares fit to the background coronal $\mathrm{B}_{\mathrm{T}}$ as described in Section~\ref{Sec:Model_background_ThomsonScattering}.  For the coronal magnetic field, we use the single power law representation that appears in \cite{Kooi:2014}:
\begin{eqnarray}
\mathbf{B}(r) = B_0 r^{-\delta} \mathbf{\hat{e}_r}\label{eq:single_B_background}
\end{eqnarray} 
where $r$ is in units of $R_\odot$ and $B_0$ and $\delta$ are taken from the model of \cite{Sakurai&Spangler:1994a}: $B_0 = 1.01$ G and $\delta = 2$.  The constant $B_0$ can be of either polarity and reverses sign at the coronal current sheet.  From Equation (1), 
the resulting expression for rotation measure, RM, is 
\begin{equation}\label{eq:RM_model_background}
\mathrm{RM} = \left[ \frac{2C_{FR} R_{\odot} N_0 B_0}{(\gamma -1) R_0^{\gamma-1}}  \right] \left(\cos^{\gamma-1}\beta_{c,1}-\cos^{\gamma-1}\beta_{c,2}\right)
\end{equation} 
where $C_{FR} \equiv e^3/2 \pi m_e^2 c^4, \gamma \equiv \alpha + \delta$, and $R_0$ is defined in Figure 1 and given in solar radii.  $\beta_{c,1}$ and $\beta_{c,2}$ give the locations of the first and second neutral lines; consequently, for lines of sight to sources such as 0900, the second crossing at $\beta_{c,2}$ serves to reduce the magnitude of the observed RM.    The sign of the rotation measure depends on the polarity of $\mathbf{B}$ for $\beta < \beta_{c,1}$ and the relation between $ \beta_{c,1}$ and $ \beta_{c,2}$:
\begin{enumerate}
\item If $|\beta_{c,1}|>|\beta_{c,2}|$: then $RM > 0$ when $B_0 > 0$ for $\beta < \beta_{c,1}$, otherwise $RM < 0$.
\item If $|\beta_{c,1}|<|\beta_{c,2}|$: then $RM < 0$ when $B_0 > 0$ for $\beta < \beta_{c,1}$, otherwise $RM > 0$.
\end{enumerate}
For lines of sight that only cross one neutral line, $\beta_{c,2}\equiv\pi/2$ and Equation~\eqref{eq:RM_model_background} reduces to Equation (9) in \cite{Kooi:2014}.  The expression Equation~\eqref{eq:single_B_background} is in cgs units.  For MKS units (the conventional units of rad m$^{-2}$), the number resulting from Equation~\eqref{eq:single_B_background} should be multiplied by $10^4$. 

We do not perform a least-squares fit to determine the magnitude $B_0$, but elect to use the same value, $B_0 = 1.01$ G, as above
because we only have at most two to three radioastronomical scans of 0842, 0843, and 0900 before they were occulted by CMEs; the only parameter calculated from a fit to data in Equation~\eqref{eq:RM_model_background} is $N_0$.  Consequently, the RM curve for the background corona is a {\it prediction} and not a fit.  The RM curve for the background coronal plasma given by Equation~\eqref{eq:RM_model_background} is shown as a 
dotted line in the RM(t) profile of Figures~\ref{fig:RM_Time_Series_0842_withmodel},~\ref{fig:RM_Time_Series_0900_withmodel}, and~\ref{fig:RM_Time_Series_0843_withdoublemodel}.  Table~\ref{T:background_model_parameters} gives the values for $\beta_c$ and $N_0$ determined for each source, the extrapolated plasma density at 1 a.u., $n_e$($r=1$ a.u.), and, for completeness, $B_0$.
\begin{table}[htbp]
	\centering
	\caption{Model Parameters for the Background Coronal Plasma}\label{T:background_model_parameters}
	\smallskip
	\begin{threeparttable}
		\begin{tabular}{cccccc}
		\hline\hline \noalign{\smallskip} 
		\multirow{2}{*}{Source}	&	\multirow{2}{*}{$\beta_{c,1}$}	&	\multirow{2}{*}{$\beta_{c,2}$\tnote{a}}	&	$N_0$\tnote{b}	&  	$n_e$($r=1$ a.u.)\tnote{c}		&$B_0$\tnote{d}\\
							&							&									&	($10^5$ cm$^{-3}$)&(cm$^{-3}$)& (G)\\
		\hline\noalign{\smallskip}
		0842	&	$-22\fdg4- -25\fdg4$	&	$-$				&	$3.05\pm0.25$		&	2.8		&1.01\\
		0843	&	$19\fdg0-21\fdg3$	&	$-$				&	$2.19\pm0.26$			&	2.1		&1.01\\
		0846	&	$21\fdg7-22\fdg8$	&	$-$				&	$3.10\pm0.47$			&	2.9		&1.01\\
		0900	&	$30\fdg4-30\fdg8$	&	$-44\fdg2- -52\fdg4$	&	$4.20\pm0.58$	&	4.0		&1.01\\
		\noalign{\smallskip} 
		\hline
		\end{tabular}
	\medskip
		\begin{tablenotes}
		\footnotesize
		\item[a] Only 0900 was occulted by two neutral lines.
		\item[b] Determined from a least-squares fit to the Thomson scattering brightness attributed to the background corona.
		\item[c] Determined by assuming Equation~\eqref{eq:single_nepower} holds out to $10R_\odot$, then extrapolating out to 1 a.u. with $n_e(r)\propto r^{-2}$.
		\item[d] Value taken from  \cite{Sakurai&Spangler:1994a}.
		\end{tablenotes}	
	\end{threeparttable}
\end{table}
The $N_0$ determined for 0842 and 0846 are consistent with each other; however, $N_0$ for 0843 is somewhat smaller likely because the line of sight to this source sampled the dimmest region of the corona relative to the other sources, prior to occultation by CME-1.  Similarly, $N_0$ for 0900 is somewhat larger because the source line of sight samples the edge of a bright streamer prior to occultation (e.g., see Figure~\ref{fig:CME_occult_geometry}).  To compare the $N_0$ values to the plasma density measured in situ at 1 a.u., we assume Equation~\eqref{eq:single_nepower} holds out to $10R_\odot$, then extrapolate out to 1 a.u. with $n_e(r)\propto r^{-2}$; the extrapolated plasma densities in Table~\ref{T:background_model_parameters} range from $2.1$ cm$^{-3}$ to $4.0$ cm$^{-3}$.   Over the period of August 2 through August 6, 2012, the Charge, Element, and Isotope Analysis System (CELIAS) Proton Monitor (PM) on {\it SOHO} measured a range of proton densities ($1.3-13.5$ cm$^{-3}$) with a mean value of $\sim4.6$ cm$^{-3}$.  While our values for $N_0$ in Table~\ref{T:background_model_parameters} are lower than the original model value of $N_0=1.83\times10^6$ cm$^{-3}$ used by \cite{Sakurai&Spangler:1994a}, our values are consistent with CELIAS/PM plasma density data for this period; however, this calculation is contingent on the heliocentric distance at which $n_e(r)\propto r^{-2}$.  Further discussion and comparison between data and model is deferred to Section~\ref{sec:results} below.


\subsection{Faraday Rotation through a Force-Free Flux Rope}\label{Sec:Model_FFrope_FR}

We model the CME as a cylindrically symmetric force-free flux rope with a magnetic field composed of an axial and azimuthal field \citep[e.g., see][]{Gurnett&Bhattacharjee:2006}:
\begin{equation}\label{eq:flux_rope_Bfield}
\mathbf{B} = B_{cme}\left[J_0\left(\alpha\rho\right) \mathbf{\hat{e}_z}+HJ_1\left(\alpha\rho\right) \mathbf{\hat{e}_\phi}\right]
\end{equation}
where $B_{cme}$ is the magnitude of the magnetic field along the central flux rope axis, $H$ specifies the helicity ($H = -1$ for left-handed and $H = +1$ for right-handed helicities), $J_0$ and $J_1$ are the zeroth- and first-order Bessel functions of the first kind, respectively, and the coordinates are given in axis-centered cylindrical coordinates $\left(\mathbf{\hat{e}_\rho},\mathbf{\hat{e}_\phi},\mathbf{\hat{e}_z}\right)$.  For a flux rope with radius $R_{cme}$, we define $\alpha R_{cme}\equiv2.405$, the first zero of $J_0$, to ensure that the the axial field is zero at the surface of the flux rope.  By employing Equation~\eqref{eq:flux_rope_Bfield}, we are assuming that the CME can be approximated by a cylinder on the scales of the LOS penetration of the CME and we do not account for the CME's curvature on global scales.

\begin{figure}
	\begin{center}
	\subfigure[Profile View of Flux Rope ($\phi_z=90\arcdeg$)]{
	\includegraphics[height=2.4in, trim = {35mm 30mm 20mm 30mm}, clip]{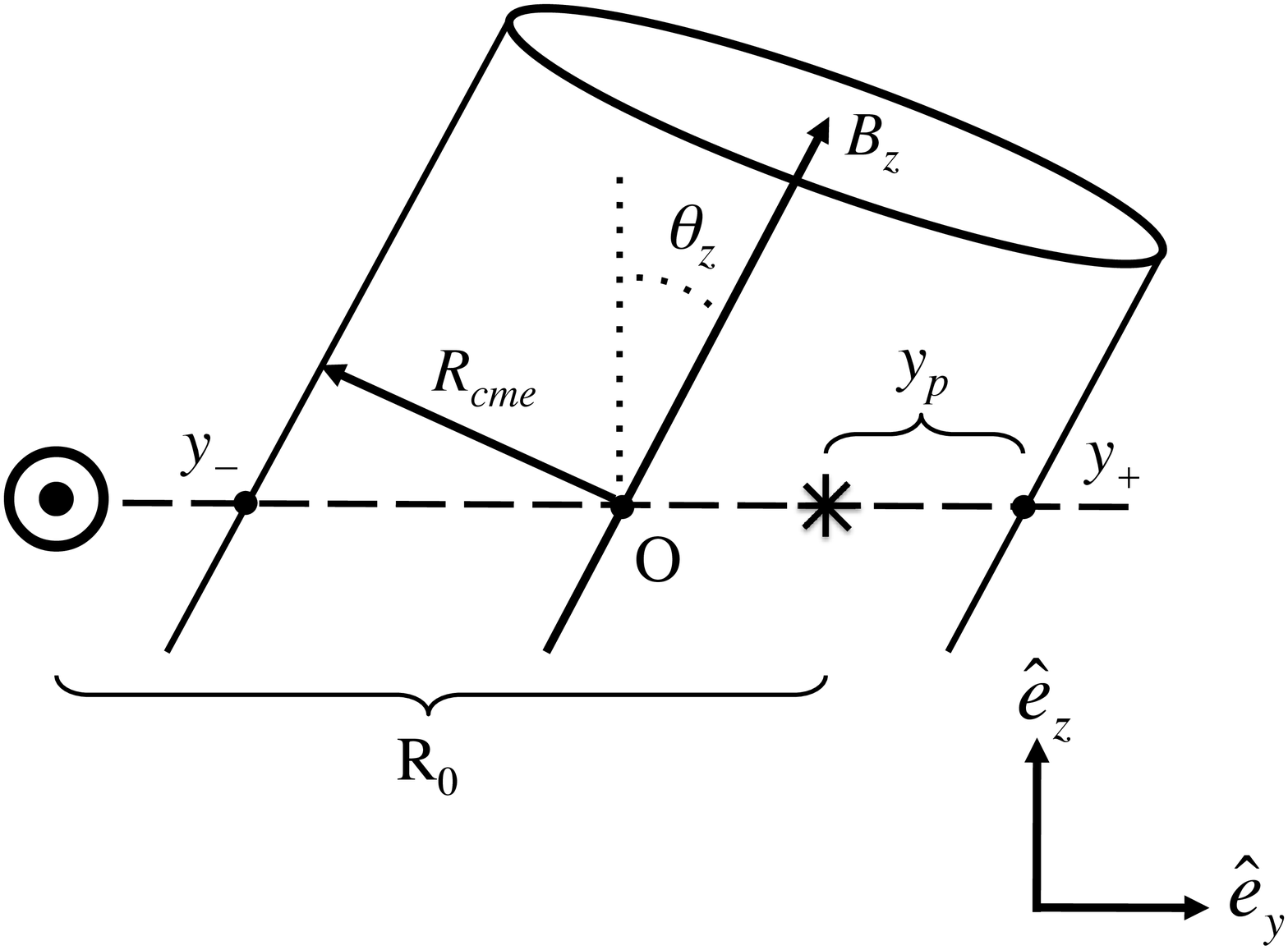}
	\label{fig:geometryprofile}
	}
	\hspace{.05in}
	\subfigure[Top-down View of Flux Rope]{
	\includegraphics[height=2.4in, trim = {20mm 30mm 20mm 40mm}, clip]{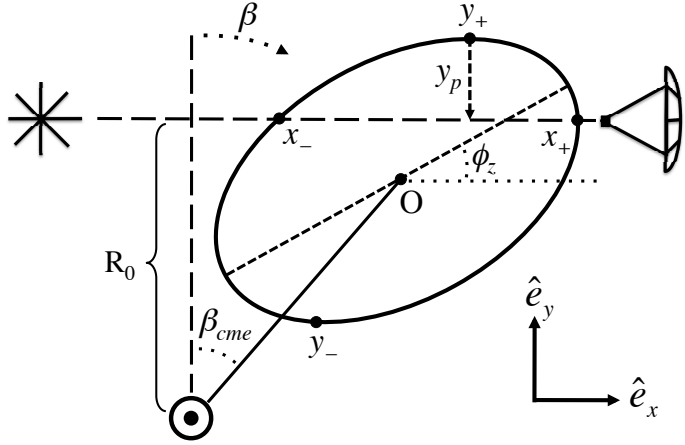}
	\label{fig:geometrytopdown}
	}	
	\caption[Illustration of the line of sight from a radio source, through a flux rope, to a radio telescope on Earth.]{\small Illustration of the LOS from a radio source, through a flux rope CME, to a radio telescope on Earth.  The LOS passes at a closest distance, or impact parameter, $R_0$.  The axial field, $B_z$ of the flux rope is rotated by $\theta_z$ with respect to the plane defined by the LOS and the Sun and the ellipse is the projection of the (tilted) flux rope on this plane; the small dashed line gives the semi-major axis of the projection.  O is the point of intersection between the plane and the central axis of the flux rope.  $\phi_z$ gives the rotation of the semi-major axis with respect to the LOS.  $x_\pm$ are the points at which the LOS intercepts the boundary of the flux rope and $y_\pm$ give the maximum extent of the flux rope as measured from the central axis in the Sun-LOS coordinate system.  $y_p$ gives the penetration depth and is the distance from the leading edge of the CME to the LOS.  The figure illustrates an idealization that will be employed in this paper, which is that the CME emerges from and continues to follow the coronal neutral line ($\beta_{cme} = \beta_{c}$), the solid line from the Sun to O.}
	\label{fig:geometry_cartoon_LOSplane}
	\end{center}
\end{figure}

Equation~\eqref{eq:flux_rope_Bfield} is given in the axis-centered reference frame of the CME.  Two Euler rotations are necessary to convert the axis-centered reference frame to the Sun-LOS reference frame.  Figure~\ref{fig:geometry_cartoon_LOSplane} shows an illustration of the Sun-LOS reference frame: in cartesian coordinates, the unit vectors $\mathbf{\hat{e}_x}$ and $\mathbf{\hat{e}_y}$ lie in the plane defined by the line of sight and the Sun (Figure~\ref{fig:geometrytopdown}) and $\mathbf{\hat{e}_z}$ is perpendicular to this plane (Figure~\ref{fig:geometryprofile}), with the origin, O, centered at the point where the central axis of the CME intersects this plane.  Figure~\ref{fig:geometry_cartoon_LOSplane} also defines the three angles that are important in determining the line of sight magnetic field component: $\theta_z$ is the angle that the axial magnetic field, $B_z$, makes with respect to the Sun-LOS plane and is defined as positive for a rotation toward the line of sight; $\phi_z$ is the angle by which the semi-major axis of the tilted flux rope has been rotated in the Sun-LOS plane; and $\beta_{cme}$ is the angle at which the flux rope was ejected from the Sun.  A flux rope with $\theta_z=0\arcdeg$ is oriented perpendicular to the Sun-LOS plane and the axial field contribution to Faraday rotation will be zero.  Similarly, a flux rope with $\theta_z=90\arcdeg$ and $\phi_z=0\arcdeg$ has an axial field aligned with the line of sight and the azimuthal contribution will be zero.

Also in Figure~\ref{fig:geometry_cartoon_LOSplane} are the limits $x_\pm$ and $y_\pm$.  $x_+$ and $x_-$ are, respectively, the points closest to and furthest from the observer at which the line of sight intercepts the boundary of the flux rope.  $y_+$ and $y_-$ are the points at which the line of sight first enters and finally exits the flux rope, respectively.  Finally, $R_0$ is the impact parameter, $R_{cme}$ is the radius of the flux rope, and $y_p$ gives the penetration depth and is the distance from the leading edge of the CME to the line of sight; $R_{cme}$ and $y_p$ are measured using the LASCO/C3 images as discussed in Section~\ref{Sec:Imaging_whitelight}.

Substituting Equation~\eqref{eq:flux_rope_Bfield} into Equation (1) 
and making the necessary rotations gives
\begin{equation}\label{eq:flux_rope_FR_equation_full}
\mathrm{RM}_{cme}=C_{cme}\int^{\tilde{u}_+}_{\tilde{u}_-}\left[J_0\left(\alpha R_{cme} \tilde{\rho}\right)\cos\phi_z\tan\theta_z-HJ_1\left(\alpha R_{cme}\tilde{\rho}\right) \frac{\tilde{y}}{\tilde{\rho}}\right]d\tilde{u}
\end{equation}
and 
\begin{eqnarray}
\tilde{u}_\pm &=& \frac{-\tilde{y}\sin\phi_z\cos\phi_z\sin\theta_z\tan\theta_z\pm\sqrt{1-\tilde{y}^2+\sin^2\phi_z\tan^2\theta_z}}{1+\sin^2\phi_z\tan^2\theta_z}\\\label{eq:flux_rope_FR_upm}
\tilde{\rho}^2 &=& a_1(\phi_z,\theta_z)\tilde{u}^2+a_2(\phi_z,\theta_z)\tilde{u}\tilde{y}+a_3(\phi_z,\theta_z)\tilde{y}^2\label{eq:flux_rope_FR_rho}
\end{eqnarray}
where $a_1(\phi_z,\theta_z)=1+\sin^2\phi_z\tan^2\theta_z$, $a_2(\phi_z,\theta_z)=\sin2\phi_z\sin\theta_z\tan\theta_z$, and $a_3(\phi_z,\theta_z)=\cos^2\theta_z+\cos^2\phi_z\sin^2\theta_z$ and $\phi_z\in\left[0,2\pi\right], \theta_z\in\left[-\pi/2,\pi/2\right]$.  In Equation~\eqref{eq:flux_rope_FR_equation_full}, the coefficient is $C_{cme} = C_{FR}N_{cme}B_{cme}R_{cme}$ and the integration variable is $\tilde{u}\equiv\tilde{x}\cos\theta_z$.  The variables $\tilde{x}$, $\tilde{y}$, and $\tilde{\rho}$ are dimensionless and have been scaled by $R_{cme}$.  In this calculation, we have assumed that the plasma density, $N_{cme}$, is constant through the flux rope structure to simplify analysis.  In relation to Figure~\ref{fig:geometry_cartoon_LOSplane}, $\tilde{u}_\pm=x_\pm\cos\theta_z/R_{cme}$ and $\tilde{y} = \sqrt{1+\sin^2\phi_z\tan^2\theta_z} - y_p/R_{cme}$.  The maximum and minimum values $\tilde{y}$ attains are $y_\pm/R_{cme}=\pm\sqrt{1+\sin^2\phi_z\tan^2\theta_z}$, which comes from the requirement that at the instant the line of sight is tangent to the surface of the flux tube, $x_+ = x_-$.  In measuring $R_{cme}$, $y_+$ is associated with the leading edge of the outer loop and $y_-$ is associated with the boundary between the inner cavity and the bright core.  Equation~\eqref{eq:flux_rope_FR_equation_full} reproduces Figure 3 in \cite{Liu:2007} for $\phi_z = 0$ and letting $y_p$ be determined by the CME velocity and $\beta_{cme}$.

Because we have assumed that the plasma density is constant inside the flux tube, the Thomson scattering brightness is given simply by
\begin{equation}
\mathrm{B}_{\mathrm{T}}/B_\odot = \left(\frac{3\sigma_T R_\odot N_{cme}}{64R_0}\right)\left[6\left(\beta_+-\beta_-\right)-\left(\sin2\beta_+-\sin2\beta_-\right)\right] \label{eq:thomson_scattering_CMEconstNe}
\end{equation}
where $\beta_\pm$ are the angles to $x_\pm$ in the $\beta$ coordinate defined in Figures (1) 
and~\ref{fig:geometrytopdown} and is given by
\begin{equation}
\tan\beta_\pm = \left[1 - \left(\frac{R_{cme}/R_\odot}{R_0}\right)\tilde{y}\right]\tan\beta_{cme}+\left(\frac{R_{cme}/R_\odot}{R_0}\right)\tilde{x}_\pm\label{eq:thomson_scattering_betapmdef}
\end{equation}
Between Equations~\eqref{eq:flux_rope_FR_equation_full} and~\eqref{eq:thomson_scattering_CMEconstNe}, there are six free parameters: $\beta_{cme}$, $\theta_z$, $\phi_z$, $N_{cme}$, $H$, and $B_{cme}$.  we determined these parameters for CME-1 and CME-2 using the following method:
\begin{enumerate}
\item We assume $\beta_{cme}=\beta_c$.  As discussed in Section~\ref{sec:CME_properties}, both CME-1 and CME-2 were ejected near the coronal neutral line determined from the WSO potential field source surface model and, consequently, we assume the CMEs continue to follow the neutral line out to a given source's line of sight.  
\item We calculate $\theta_z$ from the LASCO/C3 images by measuring the angle that the leading edge makes with the Sun-LOS plane (e.g., see Figure~\ref{fig:geometryprofile}).  To do this, we assume (1) the leading edge is parallel to the central axis and (2) the measured angle is not subject to significant projection effects.  The latter assumption would not be valid if $\beta_{cme}$ were large; however, $\beta_{cme}$ must be small otherwise geometric projection effects would make the three-part structure of CME-1 and CME-2 difficult to decipher (e.g., at large $\beta_{cme}$ values, CME-1 and CME-2 would become partial halo CMEs).
\item Because we have assumed that the flux rope is ejected at an angle $\beta_{cme}=\beta_c$ and we further assume the semi-major axis of the flux rope in the Sun-LOS plane will be oriented in the same direction, then $\phi_z = \pm90\arcdeg-\beta_c$ where $\pm$ refer to CMEs ejected from the western and eastern solar limbs, respectively.
\item We determined $N_{cme}$ by performing a least-squares fit of Equation~\eqref{eq:thomson_scattering_CMEconstNe} to the Thomson scattering brightness (after removing the model background coronal contribution).
\item We selected the sign for the flux rope helicity, $H$, to give the appropriate magnetic polarity for the line of sight magnetic field geometry required by the rotation measure time series.
\item We determined $B_{cme}$ by performing a least-squares fit of Equation~\eqref{eq:flux_rope_FR_equation_full} to the rotation measure time series using the previously calculated $N_{cme}$ (again, after removing the model background coronal contribution).
\end{enumerate}
In removing the model background coronal contribution, it is important to account for the region along the line of sight within the flux rope.
The same method was also applied for CME-3; however, because CME-3 does not have a three-part structure we additionally assume that $R_{cme}\approx 3R_\odot$, which is within the range of $R_{cme}$ measured for CME-1 and CME-2 (see Table~\ref{T:CME_model_parameters}).

The values determined for $\beta_{cme}$, $\theta_z$, $\phi_z$, $N_{cme}$, and $B_{cme}$ for each CME appear in Table~\ref{T:CME_model_parameters}.  The $\mathrm{B}_{\mathrm{T}}$ and RM(t) curves for the flux rope models corresponding to these values are shown as dashed lines in Figures~\ref{fig:RM_Time_Series_0842_withmodel},~\ref{fig:RM_Time_Series_0900_withmodel}, and~\ref{fig:RM_Time_Series_0843_withdoublemodel} here in Section~\ref{Sec:Model}.  Further discussion of the significance of the comparison of data and model is deferred to Section~\ref{sec:results} below.


\section{Comparison of Observations with Coronal and CME Models}\label{sec:results}

We begin by demonstrating our background coronal model's capability to reproduce observations of 0846 because our ability to estimate the background corona is crucial for correctly applying the CME model and interpreting the CME data.  We then employ a single flux rope model to reproduce observations of 0842 and 0900.  Finally, we use a two flux rope model (corresponding to CME-1 and CME-2) to reproduce observations of 0843.


\subsection{0846: Coronal Occultation Only}\label{sec:results_RM0846}

\begin{figure}[htb!]
	\begin{center}
	\includegraphics[width=4.5in, trim = {10mm 0mm 10mm 10mm}, clip]{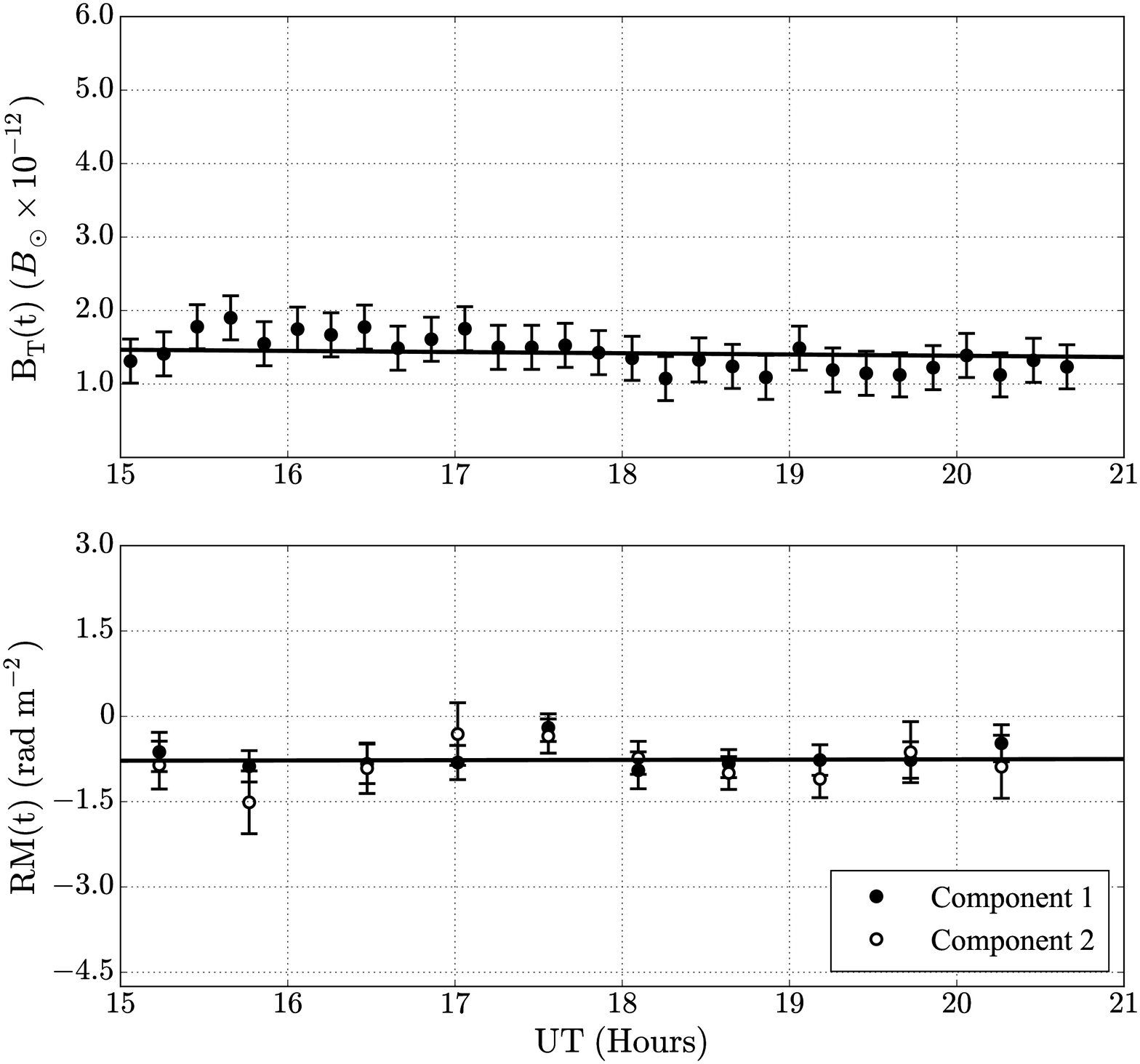}
	\caption[Thomson Scattering and coronal Faraday rotation to 0846 on 2012 August 2: occultation by the corona only.]{\small Thomson scattering brightness ({\it top}) and coronal RM(t) ({\it bottom}) for 0846 on 2012 August 2.  Thomson brightness is given for one LOS to the target source center; RM(t) is given for the LOS for Component 1 and Component 2.  Each brightness measurement is taken from one LASCO/C3 image.  Each RM measurement is determined from all 7 bandpasses for a given scan ($\sim3$ minutes duration).  Solar impact parameter $R_0$ increases from $11.1 R_\odot$ at 15:06 UT to $11.4 R_\odot$ at 21:11 UT.  The superposed curves are fits determined from the coronal models for $n_e$ and $\mathbf{B}$.  This source was occulted by the coronal plasma only and serves as a reference for comparison to sources occulted by CMEs.}
	 \label{fig:RM_Time_Series_0846}
	\end{center}
\end{figure}

The time series of Thomson brightness, $\mathrm{B}_{\mathrm{T}}$(t), and coronal Faraday rotation, RM($t$), to the source 0846 are shown together in Figure~\ref{fig:RM_Time_Series_0846} along with fits to the data determined from the coronal power law models for $n_e$ and $\mathbf{B}$ discussed in Section~\ref{Sec:Model_background_corona}.
The Thomson brightness diminishes slowly over the course of the observing session as the solar impact parameter for 0846 increases from $11.1 R_\odot$ at 15:06 UT to $11.4 R_\odot$ at 21:11 UT.  While fluctuations are present, the data do not deviate significantly from the fit and only range in value from $1-2\times10^{-12}B_\odot$, suggesting that no transient white-light structures occulted the line of sight.  The lack of apparent white-light structures in these data over the course of the observing session supports our assertion that 0846 was not occulted by a CME or other similarly complex plasma structures on August 2.  Therefore, 0846 demonstrates the effects from the background coronal plasma only and serves as a reference for comparison to sources occulted by CMEs.

The data for the rotation measure on a scan-by-scan basis show that the RM remained relatively constant during this observing session.  The solid points ({\it Component 1}) give the RM determined for the strongly polarized northern lobe of 0846 and the open symbols ({\it Component 2}) give the RM determined for the weaker southern lobe; the error bars represent the propagation of radiometer noise and are larger for the southern lobe because of its weaker polarized intensity (see Section~\ref{Sec:Imaging_Radio}).  The RM(t) for the northern and southern lobes are consistent with each other.  RM(t) could not be determined for the other two components, the northern hotspot and southern jet, described in Section~\ref{sec:source_properties} because their polarized intensities were too small; however, a mean RM for the whole observing session on August 2 was calculated for both components: the mean RM for the northern hotspot and southern jet were $-0.44\pm0.52$ rad m$^{-2}$ and $-0.41\pm0.98$ rad m$^{-2}$, respectively.  These are consistent with the RM(t) for the northern and southern lobes.

The RM(t) are small (e.g., $-0.95\pm0.32$ rad m$^{-2}$ and $-0.73\pm0.29$ rad m$^{-2}$ for the northern and southern lobes, respectively, at 18:06 UT).  The model RM (solid curve in Figure~\ref{fig:RM_Time_Series_0846}), which was determined from the fit for plasma density from the Thomson brightness data and a coronal magnetic field model given in detail in Section~\ref{Sec:Model_background_corona}, agrees well with the measured RM(t) over the entire observing session.  These small RM(t) can be qualitatively understood as a consequence of the geometry involved in making these measurements: 0846 is at large heliocentric distances where coronal Faraday rotation is expected to be at most on the order of a few rad m$^{-2}$.

Figure~\ref{fig:RM_Time_Series_0846} is particularly important in context here as it demonstrates our ability to model the background coronal plasma.  The model $\mathrm{B}_{\mathrm{T}}$ and RM profiles (solid lines in Figure~\ref{fig:RM_Time_Series_0846}) agree well with the measured data; suggesting that Equations~\eqref{eq:thomson_scattering_powerlaw} and~\eqref{eq:RM_model_background} are sufficient for modeling the Thomson scattering brightness and rotation measure contributions from the background corona.


\subsection{0842 and 0900: Occultation by a Single CME}\label{Sec:Model_0842}

\begin{figure}[htb!]
	\begin{center}
	\includegraphics[width=4.5in]{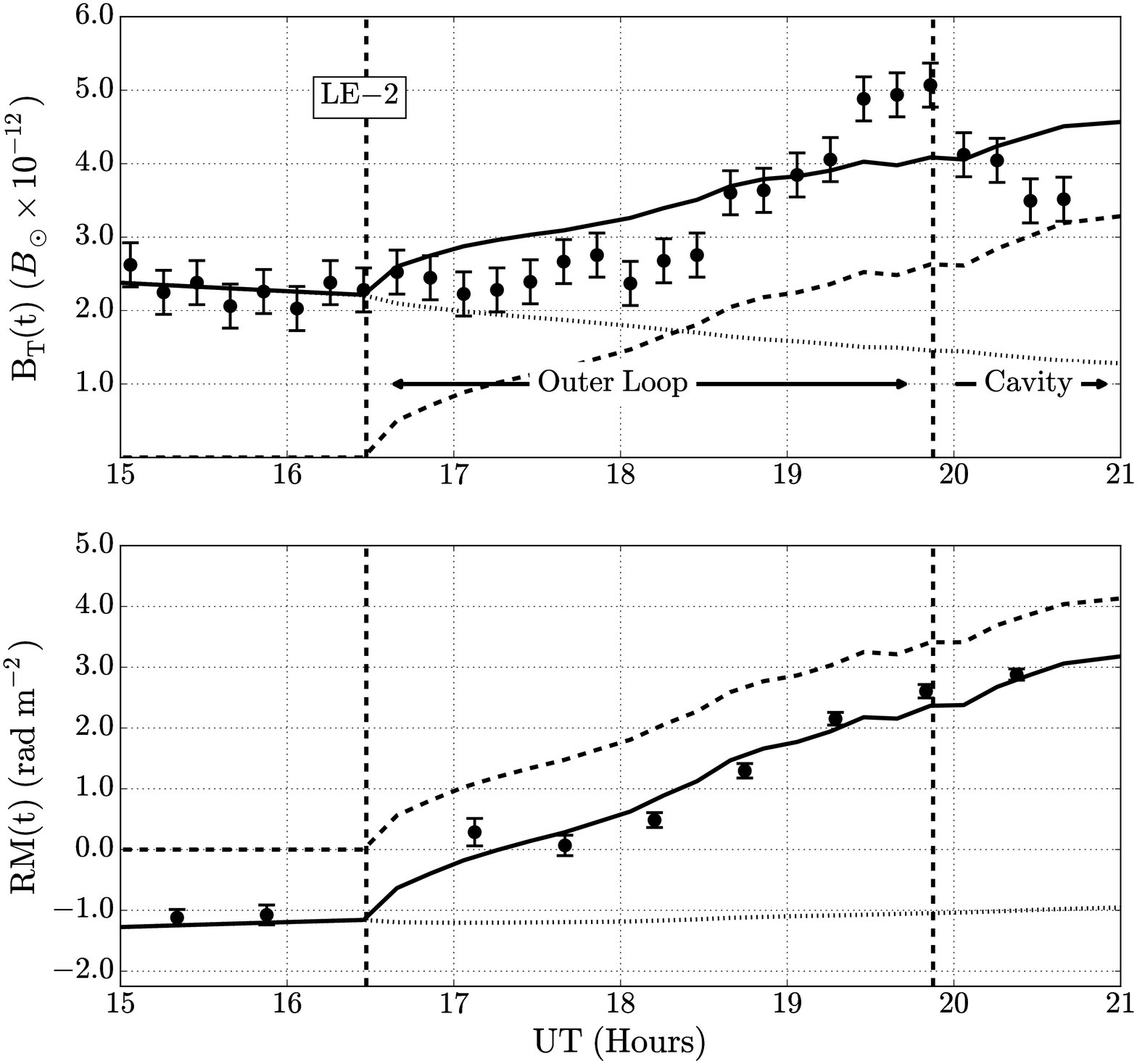}
	\caption[Thomson Scattering and coronal Faraday rotation to 0842 on 2012 August 2 including model data for a single flux rope.]{\small Thomson scattering brightness ({\it top}) and coronal RM(t) ({\it bottom}) for 0842 on 2012 August 2.  Dotted curve represents the background coronal model, dashed curve represents the single flux rope model, and solid curve represents the sum of the contributions from both models together.  Fitted parameters for the flux rope model are $N_{cme} = 6.9\pm0.5\times10^{3}$ cm$^{-3}$ and $B_{cme} = 10.4\pm0.4$ mG with helicity $H=-1$.  First vertical line (LE-2) gives the time (16:30 UT) at which 0842 was occulted by CME-2, which had the standard three-part structure.  Second vertical line gives the boundary between the outer loop and inner cavity.}
	 \label{fig:RM_Time_Series_0842_withmodel}
	\end{center}
\end{figure}

In this section, we describe the results for the two sources, 0842 and 0900, that were occulted by a single CME.  0842 is a strongly polarized point source and, thus, provides one line of sight through the plasma structure of CME-2.  Figure~\ref{fig:RM_Time_Series_0842_withmodel} shows the Thomson brightness and coronal Faraday rotation to 0842 together with the model for the background corona alone (dotted curve), the flux rope model for the CME alone (dashed curve), and the sum of the contributions from both models (solid curve).
In these fits, $\phi_z \approx90\arcdeg-\beta_{cme} = 90\arcdeg-\beta_c$ where $\beta_c$ ranges in value ($-25\fdg4\leq\beta_c\leq-22\fdg4$) over the course of the observing session as the line of sight geometry changes and $\theta_z\approx-10\arcdeg$ for the whole observing session.  The least-squares fit to $\mathrm{B}_{\mathrm{T}}$ gives $N_{cme} = 6.9\pm0.5\times10^{3}$ cm$^{-3}$ and the corresponding fit to RM(t) gives $B_{cme} = 10.4\pm0.4$ mG.  These values are summarized in Table~\ref{T:CME_model_parameters}.

\begin{table}[htbp]
	\centering
	\caption{Model Parameters for the Coronal Mass Ejections}\label{T:CME_model_parameters}
	\smallskip
	\begin{threeparttable}
		\begin{tabular}{lcccc}
		\hline\hline \noalign{\smallskip} 
		Coronal Mass Ejection	&	CME-1		&  \multicolumn{2}{c}{CME-2}		&	CME-3\\
		\hline\noalign{\smallskip}
		$R_{cme}$ ($R_\odot$)	&	$2.8-3.8$					&	\multicolumn{2}{c}{$2.5-4.0$}					&	$3.0$\tnote{a}	\\
		$\beta_{cme}$			&	$19\fdg0-21\fdg3$			&	\multicolumn{2}{c}{$-22\fdg4- -25\fdg4$}			&	$30\fdg4-30\fdg8$\\
		$\phi_z$				&	$90\arcdeg-\beta_{cme}$		&	\multicolumn{2}{c}{$90\arcdeg-\beta_{cme}$}		&	$-90\arcdeg-\beta_{cme}$\\
		$\theta_z$		&	$80\arcdeg$					&	$80\arcdeg$\tnote{a}	&	$-10\arcdeg$\tnote{b}	&	$45\arcdeg$\\
		$N_{cme}$ ($10^3$ cm$^{-3}$)	&	$21.4\pm0.6$		&	\multicolumn{2}{c}{$6.9\pm0.5$}					&	$11.2\pm0.3$\\
		$B_{cme}$ (mG)	&	$11.3\pm0.4$					&	\multicolumn{2}{c}{$10.4\pm0.4$}					&	$2.4\pm0.3$\\
		$H$				&	$+1$							&	\multicolumn{2}{c}{$-1$}						&	$-1$\\
		\noalign{\smallskip} 
		\hline
		\end{tabular}
	\medskip
		\begin{tablenotes}
		\footnotesize
		\item[a] $\theta_z$ for 0843.
		\item[b] $\theta_z$ for 0842.
		\end{tablenotes}	
	\end{threeparttable}
\end{table}

Prior to occultation by CME-2, the Thomson brightness for 0842 was $60\%$ larger than the Thomson brightness for 0846; this is because 0842 was observed at smaller impact parameters, $9.4 R_\odot-10.6 R_\odot$.  The trend is the same, though, as the Thomson brightness slowly diminishes until the leading edge (LE-2) crosses the line of sight at 16:30 UT.  0842 is first occulted by the outer loop of CME-2, represented by an initially slow increase in $\mathrm{B}_{\mathrm{T}}$(t) until 18:06 UT at which $\mathrm{B}_{\mathrm{T}}$(t) begins increasing more rapidly before reaching a maximum value of $\sim5\times10^{-12}B_\odot$.  At 20:06 UT, the line of sight begins penetrating the inner cavity of CME-2; although there is a corresponding decrease in Thomson brightness, $\mathrm{B}_{\mathrm{T}}$(t) remains $\sim2\times$ greater than the background coronal Thomson scattering model (dotted line in Figure~\ref{fig:RM_Time_Series_0842_withmodel}).

The RM(t) for 0842 also demonstrate a strong signal associated with the passage of CME-2.  Prior to occultation, the RM(t) is near $-1$ rad m$^{-2}$ and is in very good agreement with the model RM for the background corona determined in the same way as 0846 (see Section~\ref{Sec:Model_background_corona}).  After occultation by the outer loop of CME-2, the RM(t) changes sign and increases gradually to $2.60\pm0.11$ rad m$^{-2}$.  The sign change implies that the density enhancement associated with the increasing Thomson brightness profile will not be sufficient to account for the increasing RM(t); the magnetic field structure must also be fundamentally different to produce a sign change in the magnetic field component parallel to the line of sight.  Once the line of sight begins to sample the inner cavity, $\mathrm{B}_{\mathrm{T}}$(t) decreases, corresponding  to a decrease in the plasma density; however, the RM(t) increases to $2.88\pm0.09$ rad m$^{-2}$, for a total change of $+4.0$ rad m$^{-2}$ over the background coronal RM.  This implies an enhancement in the magnetic fields sampled by the line of sight.  We did observe 0842 at 16:35 UT, shortly after occultation by the leading edge in white-light LASCO/C3 data.  The measured RM, $-21.95\pm3.11$, dwarfed the values presented in Figure~\ref{fig:RM_Time_Series_0842_withmodel}; however, the Stokes $I$, $Q$, and $U$ maps for this scan are very poor in quality, having $\gtrsim20\times$ the noise of the other scan maps.

The background models in Figure~\ref{fig:RM_Time_Series_0842_withmodel} are given by Equations~\eqref{eq:thomson_scattering_powerlaw} and~\eqref{eq:RM_model_background} prior to occultation by LE-2; however, the background models in Figure~\ref{fig:RM_Time_Series_0842_withmodel} remove the contribution by the coronal $n_e$ and $\mathbf{B}$ along the section of the line of sight within the flux rope (i.e., the coronal plasma model along this section of the line of sight is replaced by the flux rope model).  The single flux rope model reproduces the general increase in Thomson scattering brightness, but does not reproduce the $\sim1\times10^{-12}B_\odot$ fluctuations present after CME-2 occults the line of sight.  These fluctuations are much larger than the fluctuations in the background coronal $\mathrm{B}_{\mathrm{T}}$(t) profile and are likely real.  The model overestimates $\mathrm{B}_{\mathrm{T}}$(t) near the beginning of the occultation and after the line of sight begins to sample the inner cavity; the model also underestimates $\mathrm{B}_{\mathrm{T}}$(t) during the peak occultation by the outer loop.  It is not surprising that the model produces a ``mean'' profile and does not reproduce the fast ramp and decay in $\mathrm{B}_{\mathrm{T}}$(t) because we have assumed that the plasma density is constant over the flux rope.

The single flux rope model reproduces the RM(t) data, both in sign and magnitude, for 0842.  The background model suffices for determining the RM values prior to occultation; after occultation, the addition of the flux rope model (with helicity $H=-1$) is necessary to reproduce the sign change from negative to positive near 17:08 UT.  For the $\theta_z$, $\phi_z$, $\beta_{cme} $, and $H$ determined for this flux rope, the line of sight geometry is such that the azimuthal component of the magnetic field dominates the flux rope contribution, providing the positive RM necessary to match the RM(t).  

The model appears to do a better job fitting the RM(t) profile and deviations from this fit appear to be less significant than the deviations in the $\mathrm{B}_{\mathrm{T}}$(t) profile.  The error in $\mathrm{B}_{\mathrm{T}}$(t) is comparable to the RMS deviations from the background coronal model for $\mathrm{B}_{\mathrm{T}}$(t) and is of the order $0.3\times10^{-12}B_\odot$.  Consequently, the deviations in the $\mathrm{B}_{\mathrm{T}}$(t) profile after occultation by CME-2 are $2-3$ times the error.  This is at most comparable to the deviations in the RM(t) profile, which should be true because the RM measurements have a smaller footprint (restoring beam) in the corona and are more sensitive to true fluctuations associated with the internal structure of the CME.

\begin{figure}[htb!]
	\begin{center}
	\includegraphics[width=4.5in]{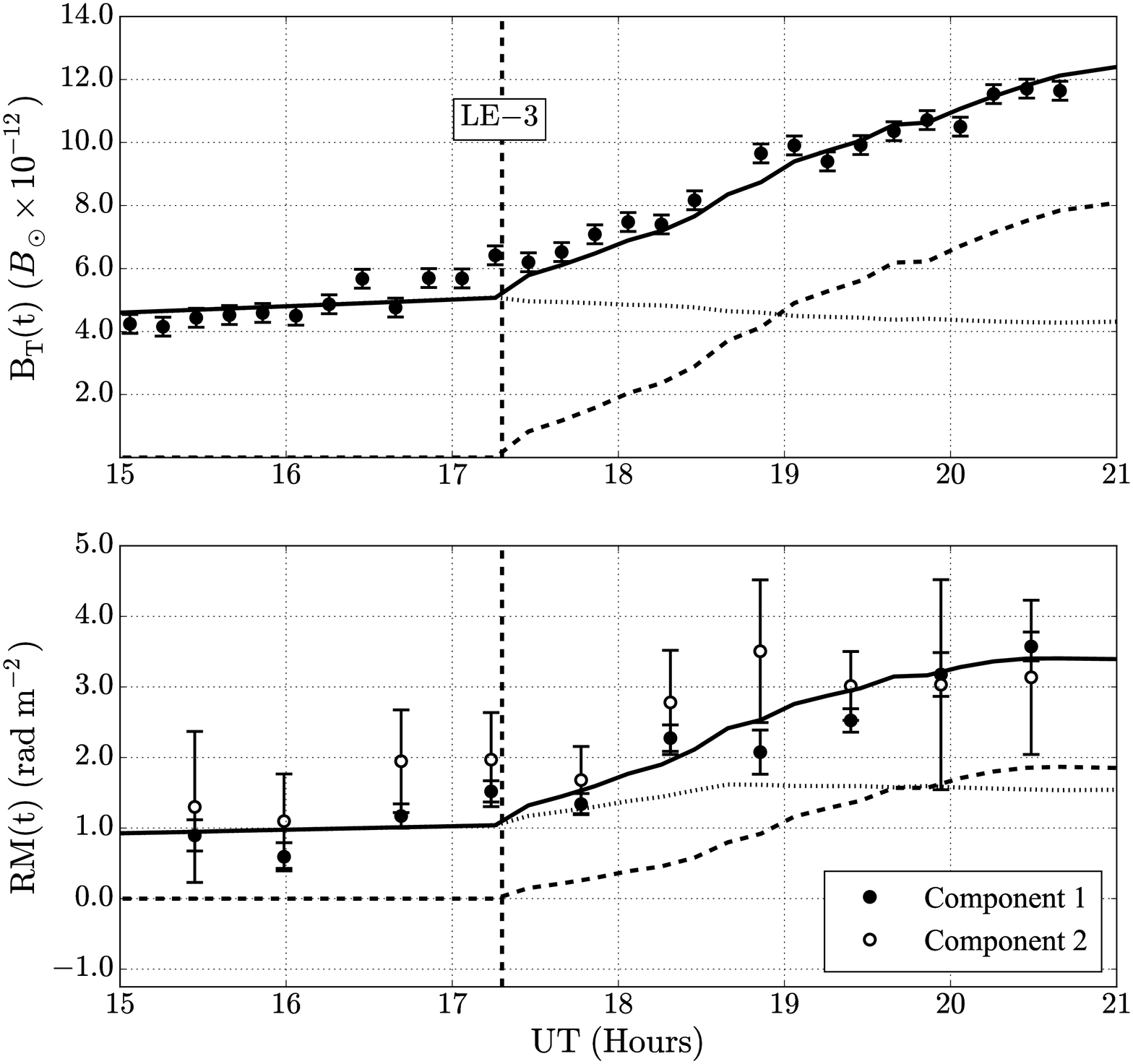}
	\caption[Thomson Scattering and coronal Faraday rotation to 0900 on 2012 August 2 including model data for a single flux rope.]{\small Thomson scattering brightness ({\it top}) and coronal RM(t) ({\it bottom}) for 0900 on 2012 August 2.  Thomson brightness is given for one LOS to the target source center; RM(t) is given for the LOS for Component 1 and Component 2.  Dotted curve represents the background coronal model, dashed curve represents the single flux rope model, and solid curve represents the sum of the contributions from both models together.  Fitted parameters for the flux rope model are $N_{cme} = 11.2\pm0.3\times10^{3}$ cm$^{-3}$ and $B_{cme} = 2.4\pm0.3$ mG with helicity $H=-1$.  First vertical line (LE-3) gives the time (17:18 UT) at which 0900 was occulted by CME-3, which had the jet-like structure of a narrow CME.}
	 \label{fig:RM_Time_Series_0900_withmodel}
	\end{center}
\end{figure}

While 0842 was occulted by a CME with the standard three-part structure, 0900 was occulted by the narrow, jet-like CME-3.  Unlike 0842, 0900 is an extended radio source (see Figure~\ref{fig:Source_Maps}) and provides multiple lines of sight through CME-3.  We report the RM data for the component in the northern and southern lobes with the strongest polarization in Figure~\ref{fig:RM_Time_Series_0900_withmodel}, along with the Thomson brightness profile, together with the model for the background corona alone (dotted curve), the flux rope model for the CME alone (dashed curve), and the sum of the contributions from both models (solid curve).
This source was occulted by two neutral lines, one on the Earth-side and one on the far-side of the Sun.  We chose $\phi_z \approx-90\arcdeg-\beta_{cme} = -90\arcdeg-\beta_c$ corresponding to the neutral line on the Earth-side of the Sun ($30\fdg4\leq\beta_c\leq30\fdg8$) because CME-3's initiation point was on the Earth-side of the Sun.  We approximated $\theta_z\approx45\arcdeg$ for the whole observing session; however, $\theta_z$ for CME-3 was not as clearly defined as it was for CME-2.  The least-squares fits to $\mathrm{B}_{\mathrm{T}}$(t) and RM(t) give $N_{cme} = 11.2\pm0.3\times10^{3}$ cm$^{-3}$ and $B_{cme} = 2.4\pm0.3$ mG.  These parameters are summarized in Table~\ref{T:CME_model_parameters}.

Of the sources discussed in this paper, 0900 had the smallest impact parameters, ranging from $8.6 R_\odot$ near the beginning of the observing period to $8.0 R_\odot$ near the end.  It is therefore no surprise that the Thomson brightness associated with the background corona is largest for this source.  Further, the general trend of the background $\mathrm{B}_{\mathrm{T}}$(t) and RM(t) to increase slowly over time is a result of the slow decrease in the impact parameter.  After the leading edge, LE-3, of CME-3 occults 0900, the Thomson brightness increases at a faster rate, approaching $\sim12\times10^{-12}B_\odot$ which is twice the predicted coronal value of $\sim6\times10^{-12}B_\odot$ (dotted line in Figure~\ref{fig:RM_Time_Series_0900_withmodel}).

The RM transient signal in the RM(t) for 0900 is not as strong as the one present in the RM(t) for 0842.  The RM is $\sim+1$ rad m$^{-2}$ at the beginning of the observing period and is in good agreement with the model for the background corona.  0900 is the only source presented in this paper that has $RM>0$ for the background corona; this is because the line of sight samples a different region of the corona on the opposite side of the Sun (see Figure~\ref{fig:Source_Maps}).  In particular, the line of sight to 0900 crosses two magnetic neutral lines, not just one as is the case for the other three sources (see Section~\ref{Sec:Model_background_Faradayrotation}).  

The difference between the RM(t) before and after occultation by CME-3 is subtle and manifests as a small increase in the rate of increasing RM; there is no sign change (as is the case in Figure~\ref{fig:RM_Time_Series_0842_withmodel}) and the total change in RM over the whole session is $\sim2.7$ rad m$^{-2}$.  This increase, though small, is detected by the strongest component in the southern lobe, {\it Component 1} in Figure~\ref{fig:RM_Time_Series_0900_withmodel}.  This southern component has the strongest polarized intensity ($P = 18.48\pm0.14$ mJy beam$^{-1}$) for this source and, therefore, small error bars.  The detection is not obvious in the RM(t) to the northern component due to its small polarized intensity ($P = 7.12\pm0.14$ mJy beam$^{-1}$) and correspondingly large error bars; however, the RM(t) to the northern component is consistent with the southern component.  Without the additional, independent data provided by the LASCO/C3, COR2-A, and COR2-B instruments, it would be difficult to interpret this RM(t) as a coronal transient.

As with Figure~\ref{fig:RM_Time_Series_0842_withmodel}, the background coronal models remove the contribution by the coronal $n_e$ and $\mathbf{B}$ along the section of the line of sight within the flux rope.  Consequently, prior to occultation by LE-3, the background model takes the same value as Equations~\eqref{eq:thomson_scattering_powerlaw} and~\eqref{eq:RM_model_background} and, after occultation by LE-3, the background coronal model values deviate.  The background model $\mathrm{B}_{\mathrm{T}}$ decreases because we have effectively removed a small fraction of the sum over plasma density. The background model RM {\it increases} as a consequence of the geometry of the line of sight: the line of sight magnetic field is negative over the majority of the line of sight removed to account for the presence of the flux rope and, therefore, a negative RM is removed, resulting in a background model RM with a larger positive magnitude.

The single flux rope model satisfactorily reproduces the general trends in both the Thomson scattering brightness and the rotation measure profiles.  Deviations from the model $\mathrm{B}_{\mathrm{T}}$(t) after occultation by CME-3 are similar to the deviations prior to occultation and are likely representative of the uncertainty in measuring $\mathrm{B}_{\mathrm{T}}$(t) and not of significant deviations from the model.  The model likely provides a better fit in this case than for the $\mathrm{B}_{\mathrm{T}}$(t) profile of 0842 because of CME-3's jet-like appearance: the line of sight is not obviously occulted by a bright outer loop and then a dark inner cavity (as is the case for 0842), it is only occulted by a bright jet-like outflow of plasma.

Similar to 0842, the model RM agrees well with the RM(t) data and there are no significant deviations, especially in the RM(t) for the strongly polarized southern lobe; however, $B_{cme}$ is smaller, largely because the differences between the pre- and post-occultation magnitudes in the profiles for 0900 are smaller than they are for 0842.  Also, the model RM for 0900 is insensitive to the parameter $\theta_z$; letting $\theta_z$ range in value from $0\arcdeg$ to $80\arcdeg$ changes $N_{cme}$ and $B_{cme}$ by less than a factor of two.  The azimuthal magnetic field dominates regardless of $\theta_z$ because the measured penetration depth, $y_p$, for CME-3 is small.

The agreement between the model and the measured RM(t) for 0900 is important for two reasons.  First, like 0846, this demonstrates our ability to accurately model the effects of the background coronal plasma; however, 0846, was only occulted by one neutral line and 0900 is occulted by two neutral lines.  If we had not accounted for the second neutral line crossing in Equation~\eqref{eq:RM_model_background}, the model value for the background coronal RM would more than double because of the line of sight geometry, producing a large discrepancy between model and measurement.  The second important feature is that our background coronal model correctly predicts $RM>0$ for 0900 and $RM<0$ for the other sources, suggesting that our background coronal models do not have a systematic bias towards negative rotation measures.

\subsection{0843: Occultation by Two CMEs}\label{Sec:Model_0843}

In this section, we describe the results for 0843, which was occulted by the outer loops of two CMEs on 2012 August 2.  \cite{Kooi:2016} demonstrated that a single flux rope model is not sufficient for reproducing the observed Thomson brightness and coronal Faraday rotation data because such a model overestimates both the Thomson scattering brightness and the rotation measure time series beginning after 19:00 UT.  The observed RM(t), in particular, diverges significantly from a single flux rope model.  The inability of a single flux rope model to reproduce the results of our observations suggests that we must account for both CMEs that occulted 0843.

To model the effect of two flux ropes occulting the line of sight to 0843, we need to determine $N_{cme}$ and $B_{cme}$ for both CMEs that occulted the line of sight: CME-1 and CME-2.  For CME-1, in performing the least-squares fit to $\mathrm{B}_{\mathrm{T}}$ and RM(t), we only fit to the data between 15:42 UT and 18:30 UT while the line of sight was only occulted by CME-1 alone; the solutions give $N_{cme} = 21.4\pm0.6\times10^{3}$ cm$^{-3}$ and $B_{cme} = 11.3\pm0.4$ mG with helicity $H=+1$.  These values are summarized in Table~\ref{T:CME_model_parameters}.
Fortunately, CME-2 also occulted 0842; consequently, we use the plasma density and axial magnetic field strength determined from the independent observations of 0842 to model CME-2: $N_{cme} = 6.9\pm0.5\times10^{3}$ cm$^{-3}$ and $B_{cme} = 10.4\pm0.4$ mG with helicity $H=-1$.

In fitting the data for CME-1, $\phi_z \approx90\arcdeg-\beta_{cme} = 90\arcdeg-\beta_c$ where $\beta_c$ ranges in value ($19\fdg0\leq\beta_c\leq21\fdg3$) over the course of the observing session as the line of sight geometry changes.  The parameter $\theta_z$ did not vary much in the cases of 0842 and 0900 because the lines of sight for those two sources penetrated CME-2 and CME-3 and progressed deeper into these CMEs; however, the line of sight to 0843 penetrates the outer loop of CME-1, traces a chord through the outer loop, and exits the backside (e.g., Figure~\ref{fig:CME_occult_geometry}).  Similarly, the line of sight samples the outer loop of CME-2, but does not pierce into the inner cavity region.  As a consequence, the orientation of the leading edge to the Sun-LOS plane evolves over the course of the observations, ranging in value from $\theta_z\approx75\arcdeg$ to $\theta_z\approx-75\arcdeg$.  For simplicity, the model results we present here use a constant value $\theta_z\approx80\arcdeg$ for CME-1.

Although the line of sight to 0843 only crosses one neutral line, which we have associated with $\beta_{cme}$ for CME-1, we assume that CME-2 crosses the line of sight at approximately the same angle as 0842: $\beta_{cme}\approx-24\arcdeg$ for CME-2.  We also assume $\phi_z = 90\arcdeg-\beta_{cme}$, applying the appropriate $\beta_{cme}$ for CME-1 and CME-2.  The only flux rope model parameter we change for CME-2 is setting $\theta_z\approx80\arcdeg$ to approximate the observations of the leading edge LE-2.

\begin{figure}[htb!]
	\begin{center}
	\includegraphics[width=4.5in]{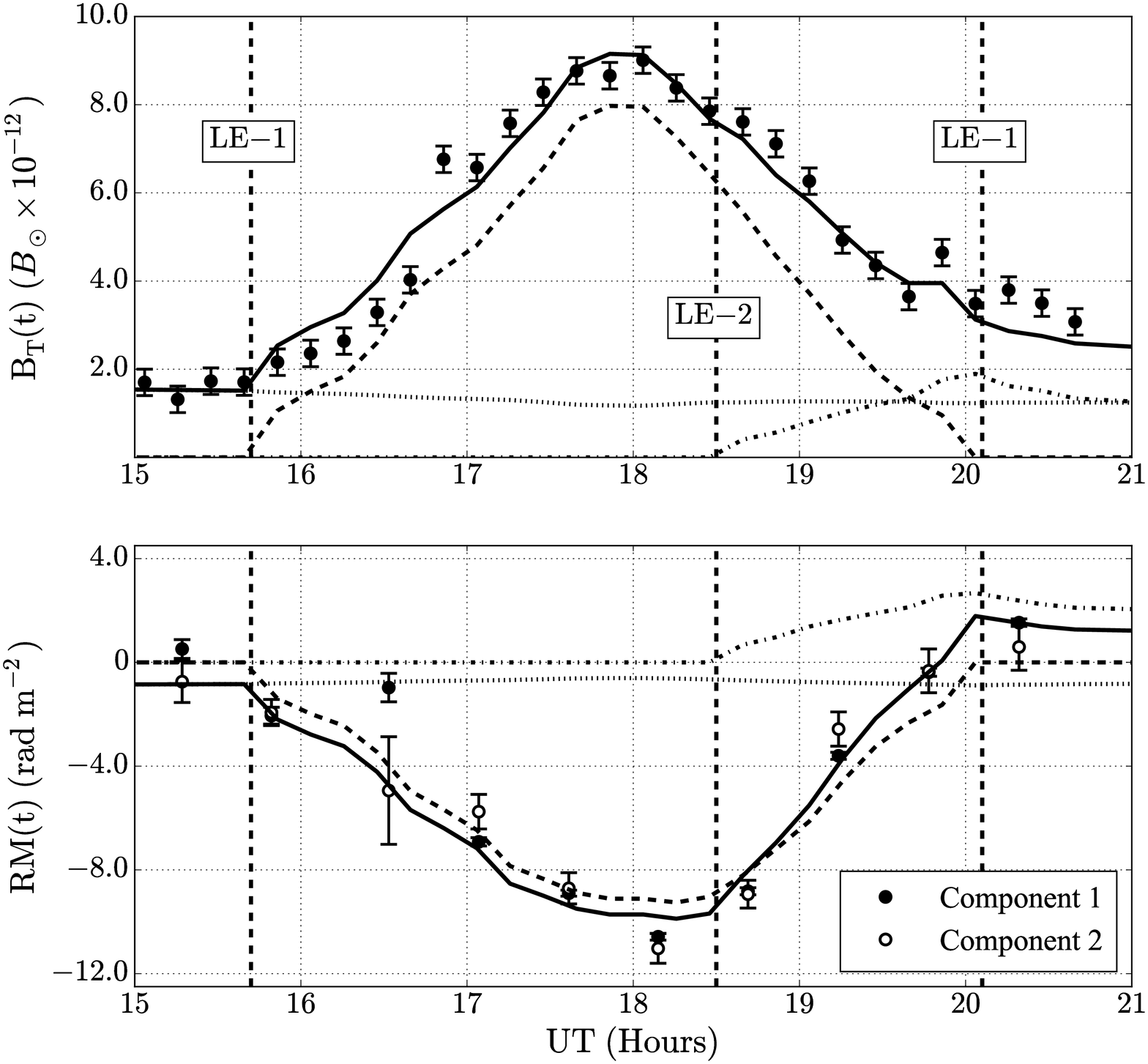}
	\caption[Thomson Scattering and coronal Faraday rotation to 0843 on 2012 August 2 including model data for two flux ropes.]{\small Thomson scattering brightness ({\it top}) and coronal RM(t) ({\it bottom}) for 0843 on 2012 August 2.  Thomson brightness is given for one LOS to the target source center; RM(t) is given for the LOS for Component 1 and Component 2.  Dotted curve represents the background coronal model, dashed curve represents the flux rope model for CME-1, dashed-dotted curve represents the flux rope model for CME-2, and solid curve represents the sum of the contributions from all models together.  Fitted parameters for the flux rope model associated with CME-1 are $N_{cme} = 21.4\pm0.6\times10^{3}$ cm$^{-3}$ and $B_{cme} = 11.3\pm0.4$ mG with helicity $H=+1$. Fitted parameters for the second flux rope model are taken directly from the fit to data for 0842. First and third vertical lines (LE-1) give the times (15:42 UT \& 20:06 UT, respectively) at which occultation by CME-1 begins and ends, respectively.  The second vertical line (LE-2) gives the time (18:30 UT) at which occultation by CME-2 begins.  Both CMEs had the standard three-part structure.}
	 \label{fig:RM_Time_Series_0843_withdoublemodel}
	\end{center}
\end{figure}

Figure~\ref{fig:RM_Time_Series_0843_withdoublemodel} shows the results of the two flux rope model along with the  Thomson brightness and coronal Faraday rotation data for 0843.  The models are as follows: the background corona alone (dotted curve), the flux rope model for CME-1 alone (dashed curve), the flux rope model for CME-2 alone (dashed-dotted curve), and the sum of the contributions from all models (solid curve).  
It is important to emphasize that the model sum (solid line in Figure~\ref{fig:RM_Time_Series_0843_withdoublemodel}) represents a fit to the observed data up to 18:30 UT.  After 18:30 UT, the model represents a \textit{prediction} based on the model data for CME-1 determined from the fit prior to 18:30 UT and the model data for CME-2 determined from the independent measurements of 0842 and the background coronal model.

Both the white-light and Faraday rotation observations for 0843 demonstrate significant transients.  The impact parameters for this source, ranging from $9.8 R_\odot$ at 15:06 UT to $10.5 R_\odot$ at 21:11 UT, are larger than those of 0900 and comparable to those of 0842; however, the transient signals measured for those two sources are much smaller by comparison.  The nominal Thomson scattering brightness from the background corona is $\sim1.5\times10^{-12}B_\odot$ at the beginning of the observing period; after occultation by the leading edge, LE-1, of CME-1 at 15:42 UT, the brightness begins increasing rapidly until peaking two hours later at $\sim9.0\times10^{-12}B_\odot$, six times the value associated with the background corona.  $\mathrm{B}_{\mathrm{T}}$(t) begins decreasing after 18:06 UT and continues to do so after occultation by the leading edge, LE-2, of CME-2; this is the same CME that occults the line of sight to 0842.  Near 20:06 UT, close to the end of the observing period, CME-1 ceases to occult the line of sight; however, because CME-2 continues to occult 0842, $\mathrm{B}_{\mathrm{T}}$(t) does not return to the nominal background value, but asymptotes near $3.5\times10^{-12}B_\odot$, about twice the background model.

The RM(t) profile for 0843 has a ``V'' shape trend, beginning near $-1$ rad m$^{-2}$, peaking near $-11$ rad m$^{-2}$ (quite large for these heliocentric distances), and approaching $+1.5$ rad m$^{-2}$ at the end of observations.  The peak RM during this period ($-10.58\pm0.13$ rad m$^{-2}$ and $-11.03\pm0.57$ rad m$^{-2}$ for the southern and northern components, respectively) is $>10\times$ the coronal contribution predicted as in Section~\ref{Sec:Model_background_Faradayrotation} and is correlated in time with the peak in $\mathrm{B}_{\mathrm{T}}$(t).  The $\mathrm{B}_{\mathrm{T}}$(t) ``only'' increased to $6\times$ the coronal contribution, suggesting that the enhancement in plasma density necessary to increase $\mathrm{B}_{\mathrm{T}}$(t) is not sufficient to account for the considerable increase detected in RM(t), and an enhancement in the magnetic field components along the line of sight is required.  After peaking at 18:06 UT, the RM(t) for both components decrease, approaching the background coronal value at 19:45 UT near the end of occultation by CME-1.  An interesting feature is that the rate of this decrease in magnitude ($d\mathrm{RM}/dt\approx6.6$ rad m$^{-2}$ hr$^{-1}$) is greater than the rate of increase ($d\mathrm{RM}/dt\approx-3.6$ rad m$^{-2}$ hr$^{-1}$) earlier in the session.  The last scan of 0843 suggests that RM(t)$>0$ by the end of the observing session: the measured RM for the strong southern and weaker northern components were $+1.53\pm0.14$ rad m$^{-2}$ and $+0.36\pm0.84$ rad m$^{-2}$.

The two flux rope model is able to reproduce the observational results of both $\mathrm{B}_{\mathrm{T}}$(t) and RM(t).  Single flux rope models overestimate $\mathrm{B}_{\mathrm{T}}$(t) near the end of the observations, during the slow decrease in $\mathrm{B}_{\mathrm{T}}$(t) after 18:30 UT because the contribution to the penetration depth which is associated with the lower density CME-2 in the two flux rope model is associated with the higher density CME-1 in single flux rope models.  The slow decrease in $\mathrm{B}_{\mathrm{T}}$(t) is well modeled as the contributions from the diminishing and increasing brightness profiles associated with the passage of CME-1 and CME-2, respectively, in the two flux rope model.

The real strength in the two flux rope model lies in its ability to represent the RM(t) for 0843.  The flux rope model for CME-1 is consistent with the data prior to occultation by CME-2; it gives the sign and magnitude for the RM(t) with the exception that it underestimates the peak RM by $\lesssim-1$ rad m$^{-2}$.  After the second occultation, the two flux rope model continues to successfully reproduce the data.  Again, we emphasize here that the model data after 18:30 UT is the \textit{prediction} determined from the two flux rope model; it is not a fit to the observed data.  Two striking features of this model are (1) it fits the fast slope, $d\mathrm{RM}/dt\approx6.6$ rad m$^{-2}$ hr$^{-1}$, and (2) it predicts $\mathrm{RM}>0$ at the end of the observations.  These two features result from the opposing helicities of CME-1 and CME-2.  CME-1 has a helicity $H=+1$, as determined from observations of 0843 prior to 18:30 UT, and CME-2 has a helicity $H=-1$, as determined from the independent observations of 0842.  The azimuthal magnetic field contributions to the RM(t) from CME-1 and CME-2 (the dashed and dashed-dotted lines in Figure~\ref{fig:RM_Time_Series_0843_withdoublemodel}, respectively) are negative and positive, respectively.  From 18:30 UT to 20:06 UT, the net effect gives the fast slope in RM(t) and, after CME-1 no longer occults 0843 near 20:06 UT, positive RM at the end of the observing session.

These RM data show a key feature that demonstrates an advantage of observing with extragalactic radio sources over pulsars or spacecraft transmitters: 0843 provides two closely-spaced lines of sight with strong linear polarization through CME-1 and CME-2.  The lines of sight to the stronger southern component and the northern component ({\it Component 1} and {\it Component 2} in Figure~\ref{fig:RM_Time_Series_0843_withdoublemodel}) are very close ($7\farcs8$, or 5,700 km in the corona, which is about twice the FWHM diameter of the synthesized beam) and, therefore, sample approximately similar regions of plasma.  The strong agreement between the RM(t) for both lines of sight gives confidence that this large coronal transient is real.  Another key feature of these data is their demonstration of the insight gained by employing white-light measurements from multiple vantage points.  LASCO/C3 white-light images give a clear view of the propagation of CME-1 (and the corresponding leading edge, LE-1); however, CME-2 is hard to discern in these images because it appears in the background, behind CME-1.  The leading edge and structure of CME-2 is clear, though, in COR2-A white-light images.  It is only with both sets of images that we are able to track the leading edges of CME-1 and CME-2 as their outer loops occult 0843, allowing us to employ a two flux rope model.

\section{Discussion}\label{Sec:discussion}

The $N_{cme}$ and $B_{cme}$ values determined for the flux rope models of CME-1, CME-2, and CME-3 represent an enhancement over the measured background values for the corona.  The single power law functions for $n_e$ and $\mathbf{B}$ given in Equations~\eqref{eq:single_nepower} and~\eqref{eq:single_B_background} can be evaluated at the location where the neutral line crosses the line of sight to provide an estimate of the local plasma density and magnetic field strength expected for the region occulted by a CME.  Using $R_0\approx10.2R_\odot$ and $\beta_c\approx20\arcdeg$ in the case of 0843 gives $n_e\approx0.8\times10^3$ cm$^{-3}$ and $|\mathbf{B}|\approx8.6$ mG; from Table~\ref{T:CME_model_parameters}, $N_{cme}\approx21.4\times10^3$ cm$^{-3}$ and $B_{cme}\approx11.3$ mG for CME-1, suggesting an increase in the local plasma density and magnetic field strength by factors of $\sim27$ and $\sim1.3$, respectively.  Similarly, $R_0\approx10.0R_\odot$ and $\beta_c\approx-24\arcdeg$ in the case of 0842 gives $n_e\approx1.1\times10^3$ cm$^{-3}$ and $|\mathbf{B}|\approx8.4$ mG; from Table~\ref{T:CME_model_parameters}, $N_{cme}\approx6.9\times10^3$ cm$^{-3}$ and $B_{cme}\approx10.4$ mG for CME-2, suggesting an increase in the local plasma density and magnetic field strength by factors of $\sim6.3$ and $\sim1.2$, respectively.  

Observations of 0900, however, only suggest an enhancement in the plasma density.  Using $R_0\approx8.3R_\odot$ and $\beta_c\approx30\arcdeg$ gives $n_e\approx2.0\times10^3$ cm$^{-3}$ and $|\mathbf{B}|\approx11.0$ mG; from Table~\ref{T:CME_model_parameters}, $N_{cme}\approx11.2\times10^3$ cm$^{-3}$ and $B_{cme}\approx2.4$ mG for CME-3, suggesting an increase in the local plasma density by a factor of $\sim5.6$ and a decrease in the local magnetic field strength by a factor of $\sim5$.  In making this comparison, though, it is important to distinguish between the observed structures of CME-1 and CME-2 and the structure of CME-3.  CME-1 and CME-2 both have the classic three-part structure and, therefore, it is much easier to apply and evaluate the flux rope model for these two CMEs; CME-3 has a jet-like structure and, whether this is due to geometrical projection effects of because it is a true ``narrow'' CME, the flux rope model is more difficult to constrain for this structure.  The density enhancement should also be compared to the original plasma density power law model as it appears in \cite{Sakurai&Spangler:1994a}, namely using $N_0=1.83\times10^6$ cm$^{-3}$, which is an order of magnitude larger than the $N_0$ determined for each line of sight.  Evaluating Equation~\eqref{eq:single_nepower} using this value for $N_0$ as before, we find a more modest increase over the background plasma density by a factor of $3.3$, $1.1$, and $1.3$ for CME-1, CME-2, and CME-3, respectively.  

One of the striking features of our results is our ability to represent the background coronal contribution to the observed Faraday rotation using simple single power law models for the plasma density and magnetic field.  Several models employ two or three power law terms for the plasma density \citep[e.g., see][]{Patzold:1987} or employ different model parameters depending on the region of the corona that is sampled \citep[e.g., see][]{Saito:1977}.  Similarly, the magnetic field is often represented by a dual power law in $r$, such as the sum of a dipole ($\propto r^{-3}$) and interplanetary magnetic field term ($\propto r^{-2}$) \citep{Patzold:1987,Mancuso&Spangler:2000,Kooi:2014}.  The exact form of the power laws assumed in Equations~\eqref{eq:single_nepower} and~\eqref{eq:single_B_background} should not be crucial for the results presented here because different functional forms give similar values for the narrow range of heliocentric distances ($8.0-11.4 R_\odot$) characteristic of our observations.  

Another simplification we made in modeling the background corona was assuming the coronal current sheet can be expressed as an infinitely thin magnetic neutral line where the polarity of the radial field reverses.  Both \cite{Mancuso&Spangler:2000} and \cite{Kooi:2014} found that accounting for the finite thickness and higher density of this current sheet provided better agreement with the Faraday rotation they measured.  We found excellent agreement between our models and the RM(t) (prior to and after occultation by CMEs) without accounting for the thickness or increased density of the current sheet for two reasons.  First, our observations were at larger heliocentric distances than \cite{Kooi:2014} and most of the sources observed by \cite{Mancuso&Spangler:2000}; consequently, the difference between the plasma density inside and outside the current sheet as predicted by the models in, e.g., \cite{Mancuso&Spangler:2000} is small.  Second, we have assumed that the observed CMEs follow the neutral line out to the heliocentric distances at which we observed and, therefore, the plasma structure of the CME would replace the current sheet structure during CME occultation in our models.

The Faraday rotation transients associated with CME-1, CME-2, and CME-3 were smaller than those observed by \cite{Levy:1969} and \cite{Cannon:1973}.  Two of the transients observed by \cite{Levy:1969} were at comparable impact parameters ($10.9R_\odot$ and $8.6R_\odot$ on 1968 November 4 and 8, respectively); however, these transients were $\sim40\arcdeg$ in amplitude at an observing frequency of 2.292 GHz, which corresponds to RM $\sim41$ rad m$^{-2}$.  This is four times larger than the largest transient we measured, $-11$ rad m$^{-2}$.  This is not necessarily surprising because CMEs come in a range of plasma densities and magnetic field strengths.  \cite{Cannon:1973} observed Faraday rotation transients at smaller impact parameters (see Section 1.4.1) 
and one such transient displayed an inverse ``N'' shape with a magnitude $|\mathrm{RM}|\approx7.1$ rad m$^{-2}$, which is comparable to the three transients we observed.

Comparison of our data should also be made with the work of \cite{Bird:1985} in which the measured Faraday rotation transients were directly associated with the passage of CMEs seen in {\it Solwind} coronagraph images.  In this investigation, Bird et al.~calculated the weighted mean longitudinal (line of sight) component of the magnetic field, $\overline{B}_L$, associated with the observed transients:
\begin{equation}\label{eq:meanBLOS_CME}
\overline{B}_L = \frac{1}{N_t}\int_tn_t\mathbf{B_t}\cdot\mathbf{ds}
\end{equation}
where
\begin{equation}\label{eq:meanNE_CME}
N_t = \int_tn_t\mathrm{d}s
\end{equation}
and $t$ here refers to the contribution from the coronal transient.  Equations~\eqref{eq:meanBLOS_CME} and~\eqref{eq:meanNE_CME} appear as Equation (8) in \cite{Bird:1985}.  Evaluating Equations~\eqref{eq:meanBLOS_CME} and~\eqref{eq:meanNE_CME} for CME-1, CME-2, and CME-3 using the flux rope model values obtained from observations of 0842, 0843, and 0900 gives a small range in $\overline{B}_L$: $1-6$ mG.  our values compare favorably to the values reported in Table 1 of \cite{Bird:1985}, but are smaller than the maximum observed values for $\overline{B}_L$, reported as $10-25$ mG in that paper.  The values in Table 1 of \cite{Bird:1985} were also calculated for transients located at smaller impact parameters, $4.5 - 7.6 R_\odot$.

The most recent observation of a coronal Faraday rotation transient, investigated by \cite{Ingleby:2007} and \cite{Spangler&Whiting:2009}, also had a RM profile similar to 0842: the Faraday rotation, given in \cite{Spangler&Whiting:2009} in terms of degrees, at the beginning of the observing session increases slowly from $-10\arcdeg$ to $-5\arcdeg$ over three hours and then quickly increases to $+28\arcdeg$ over the remaining three hours in the observing session.  At an observational frequency of 1.465 GHz, this $26\arcdeg$ increase corresponds to $\mathrm{RM}\sim10.9$ rad m$^{-2}$ which is comparable to the peak RM measured to 0843, although the source they observed was much closer to the Sun ($R_0=6.6 R_\odot$).  While the RM(t) is similar to that of 0842, beginning negative before quickly increasing to $\mathrm{RM}>0$, the CME \cite{Spangler&Whiting:2009} observed approaching their source did not appear to occult the source in LASCO/C2 images.

In modeling the observed CMEs, we assumed a constant density profile for the flux rope structure.  An alternative model is the graduated cylindrical shell (GCS) flux rope structure of \cite{Thernisien:2006} 
which employs an asymmetric Gaussian profile that requires the electron density to peak at the outer surface of the shell (outer loop) and fall off inside the shell (inner cavity).  Applying the GCS model to 34 CMEs from 1997 to 2002, \cite{Thernisien:2006} measured peak electron densities of $42-1730\times10^3$ cm$^{-3}$; our measurements for $N_{cme}$ are smaller (Table~\ref{T:CME_model_parameters}) because assuming a constant density effectively averages the plasma density over the entire structure.  This constant density profile is sufficient for modeling the $\mathrm{B}_{\mathrm{T}}$(t) and RM(t) for 0843 and 0900 (CME-1 and CME-3, respectively) largely because the lines of sight only sampled the bright outer loop.  The constant density profile does not adequately reproduce the $\mathrm{B}_{\mathrm{T}}$(t) for 0842, though, because the line of sight samples the outer loop and inner cavity of CME-2; in this case, the GCS model would likely provide a better fit to the data.

The simple flux rope structure for the magnetic field used in this work is very similar to those employed by \cite{Liu:2007} and \cite{Jensen:2008} with the exception that we have effectively assumed an infinite axial length and both \cite{Liu:2007} and \cite{Jensen:2008} place restrictions on this length.  We did not place restrictions on this length because we had restrictions on the geometric parameters $\theta_z$ and $\beta_{cme}$ and required $\phi_z = \pm90\arcdeg-\beta_{cme}$ ($\pm$ referring to sources off the western or eastern limb, respectively) based on LASCO/C3, COR2-A, and COR2-B observations.  Regardless, the RM calculated by \cite{Liu:2007} for a flux rope centered at $10R_\odot$ with an assumed axial field strength of 10 mG is $\sim\pm9$ rad m$^{-2}$, which is consistent with both the RM(t) we measured for 0843 and the axial field strengths we calculated for CME-1 and CME-2.  Further, in trying to model the 1979 October 23 and 24 CMEs observed by \cite{Bird:1985} at impact parameters of approximately $7.3R_\odot$ and $5.0R_\odot$, respectively, \cite{Jensen:2008} calculated an axial field strength of $\sim10$ mG, which is also consistent with our results.

Despite the simplicity of the constant density profile and the flux rope magnetic field, the model data fit the observed $\mathrm{B}_{\mathrm{T}}$(t) and RM(t) remarkably well, particularly in the case of 0843.  This is in due in part to our ability to place constraints on the model using observations from LASCO/C3, COR2-A, and COR2-B.  The white-light images from LASCO/C3 allow us to estimate the background coronal and CME plasma densities and the images from COR2-A and COR2-B provide additional vantage points from which we can track the leading edges of the occulting CMEs; this was the primary reason we could track the leading edges of CME-1 and CME-2 as they occulted 0843.  In modeling the data for 0843, we also made use of the independent measurements for 0842; without this additional line of sight, modeling the effects of two CMEs would have proven much more difficult.  These results underscore the power of performing Faraday rotation observations of CMEs with multiple lines of sight to multiple sources.

A significant improvement over these observations would require a set of VLA observations that are triggered in the event of a CME displaying favorable geometry (e.g., exhibiting the three-part structure and originating on or near the solar limb).  An effective trigger would be near real-time LASCO/C2 data.  An image with adequate quality to identify an emerging CME is available at a time which is, at most, one hour prior to present. These near real-time data are sufficient to detect CMEs when they are still low in the corona so that one may predict (1) the position angle of their eruption and (2) the approximate time of arrival at heliocentric distances of $5 - 20 R_{\odot}$.  For a set of triggered observations, $8-9$ strongly polarized, extended sources could be selected that are certain to be directly occulted by a CME.  In order to confirm or refute the flux rope model or distinguish between force-free and non-force-free flux rope models, it is imperative that there are several lines of sight through the three-part structure of the CME, ideally with at least one line of sight along the CME's axis of symmetry.

\section{Summary and Conclusions}\label{sec:Summary}

\begin{enumerate}
\item We performed polarimetric observations using the newly upgraded Very Large Array (VLA) of a constellation of extragalactic radio sources for six hours on August 2, August 5, and August 19, 2012, at heliocentric distances (our parameter $R_0$) ranging over $5 - 15 R_{\odot}$.  During the August 2 session, three radio sources were occulted by coronal mass ejections (CMEs): 0842, 0843, and 0900.  10 scans of $3-4$ minutes duration were made of each source at frequencies of $1-2$ GHz.  The data were reduced using the Common Astronomy Software Applications (CASA) data reduction package with the new ionospheric Faraday rotation correction algorithm that Jason Kooi and George Moellenbrock implemented.  These observations represent the first active hunt for CME Faraday rotation using the VLA.

\item In addition to our radioastronomical observations, we obtained white-light coronagraph images from the LASCO/C3 instrument aboard {\it SOHO} to determine the Thomson scattering brightness, $\mathrm{B}_{\mathrm{T}}$, along the line of sight to each source.  The $\mathrm{B}_{\mathrm{T}}$ is proportional to the electron plasma density sampled by the line of sight and provides a means to independently estimate the plasma density and determine its contribution to the observed Faraday rotation.

\item We determined the Thomson scattering time series, $\mathrm{B}_{\mathrm{T}}$(t), and rotation measure time series, RM(t), for each source occulted by a CME as well as a source, 0846, that was only occulted by the coronal plasma.  Large coronal transients that exceeded nominal coronal values were observed in both $\mathrm{B}_{\mathrm{T}}$(t) and RM(t) for 0842, 0843, and 0900 (Figures~\ref{fig:RM_Time_Series_0842_withmodel} -~\ref{fig:RM_Time_Series_0843_withdoublemodel}).  By contrast, the source that was only occulted by the corona did not demonstrate deviations from the $\mathrm{B}_{\mathrm{T}}$(t) and RM(t) expected for the background corona (Figure~\ref{fig:RM_Time_Series_0846}).

\item A single power law model was used for the background coronal plasma density and magnetic field.  This proved sufficient to reproduce the observed $\mathrm{B}_{\mathrm{T}}$(t) and RM(t) for 0846 as well as the other sources prior to occultation by a CME; however, our values for $N_0$ (Table~\ref{T:background_model_parameters}) are lower than the original model value of $N_0=1.83\times10^6$ cm$^{-3}$ used by \cite{Sakurai&Spangler:1994a} and subsequent work. The agreement between the background coronal model and data for 0900 are particularly important as they demonstrate the necessity of accounting for the line of sight crossing multiple magnetic neutral lines; further, this agreement demonstrates that our models for the background coronal RM are not systematically biased towards negative values.  The ability to properly model the background corona is crucial in identifying and measuring CME-related transients.

\item A constant density force-free flux rope embedded in the background corona was used to model the effects of the CMEs on $\mathrm{B}_{\mathrm{T}}$(t) and RM(t).  In the case of 0842, the flux rope model underestimated the peak value in $\mathrm{B}_{\mathrm{T}}$(t) and did not predict the decreasing $\mathrm{B}_{\mathrm{T}}$(t) inside the inner cavity region of the CME; however, there was satisfactory agreement between the model and the RM(t) (in particular, the model reproduces the sign change and gradual slope, Figure~\ref{fig:RM_Time_Series_0842_withmodel}).  For 0900, the single flux rope model successfully reproduces both the observed $\mathrm{B}_{\mathrm{T}}$(t) and RM(t) profiles (Figure~\ref{fig:RM_Time_Series_0900_withmodel}).

\item 0843 was occulted by two CMEs on 2012 August 2, as verified in {\it STEREO-A} COR2 images, and, therefore, the coronal transient observed in $\mathrm{B}_{\mathrm{T}}$(t) and RM(t) can not be satisfactorily modeled with a single flux rope; consequently, we modeled observations of 0843 using two flux ropes embedded in the background corona.  The introduction of a second flux rope is not merely the introduction of more free parameters, but is {\it required} to account for the second CME.  Further, we used the model parameters determined from the independent measurements of 0842 for the second CME to predict the $\mathrm{B}_{\mathrm{T}}$(t) and RM(t) resulting from the two flux rope model.  This two flux rope model successfully reproduces both the $\mathrm{B}_{\mathrm{T}}$(t) and RM(t) for 0843 (Figure~\ref{fig:RM_Time_Series_0843_withdoublemodel}).  In particular, the two flux rope model successfully replicates the appropriate slope in RM(t) before and after occultation by the second CME and predicts the observed change in sign to $\mathrm{RM}>0$ at the end of the observing session.

\item The Faraday rotation transients we measured were smaller than those observed by \cite{Levy:1969} and \cite{Cannon:1973}; however, the plasma densities ($6-22\times10^3$ cm$^{-3}$) and axial magnetic field strengths ($2-12$ mG) inferred from our models are consistent with the work of \cite{Liu:2007} and \cite{Jensen:2008}.  Further, the weighted mean line of sight component of the magnetic field calculated from our data gives $1-6$ mG, in agreement with the results of \cite{Bird:1985}.

\end{enumerate}


\acknowledgments
This work was supported at the University of Iowa by grants ATM09-56901 and AST09-07911 from the National Science Foundation and supported at the U.S. Naval Research Laboratory by the Jerome and Isabella Karle Distinguished Scholar Fellowship program.  The space-based coronal occultation image is courtesy of the LASCO/{\it SOHO} consortium.  {\it SOHO} is a project of international cooperation between ESA and NASA.  We thank George Moellenbrock of the NRAO staff for his patience and assistance in implementing ionospheric Faraday rotation corrections in CASA.


\newpage
\bibliographystyle{apj}
\bibliography{abbrev,Bibliography}

\end{document}